\documentclass[11pt,a4paper]{article}
\usepackage{jcappub}
\usepackage{epstopdf}
\usepackage{graphicx}
\usepackage{bm}
\usepackage{epstopdf}
\usepackage{mathtools}
\usepackage{natbib}
\usepackage{captcont}
\usepackage{booktabs}
\usepackage{threeparttable}

\title{Electroweak phase transition confronted with dark matter detection constraints}

\author[a,1]{Cheng-Wei Chiang}
\author[b,2]{, Da Huang}
\author[a,3]{, and Bo-Qiang Lu}

\affiliation[a]{Department of Physics, National Taiwan University, Taipei, Taiwan 10617, Republic of China}
\affiliation[b]{National Astronomical Observatories, Chinese Academy of Sciences, Beijing, 100012, China}

\emailAdd{chengwei@phys.ntu.edu.tw}
\emailAdd{dahuang@bao.ac.cn}
\emailAdd{bqlu@phys.ntu.edu.tw}

\abstract{We study the type-II first-order electroweak phase transition and dark matter (DM) phenomenology in both real and complex singlet extensions of SM.
In the real singlet extension with a $\mathbb{Z}_2$ symmetry, we show that the parameter regions favored by the phase transition suffer from strong
constraints from DM direct detection so that only a negligible fraction ($f_{X}\sim 10^{-4}-10^{-5}$) of DM composed of the real singlet scalar
can survive the LUX and XENON1T constraints.
In the complex singlet $S$ case, we impose a $CP$ symmetry $S\to S^{*}$ to the scalar potential.
The real component of $S$ can mix with SM Higgs boson while the imaginary component becomes a DM candidate due to the protection of the $CP$ symmetry.
By taking into account the current experimental constraints of invisible Higgs decays, Higgs signal strength measurements,
and dark matter detections, we find that there exists a large parameter space for the type-II electroweak phase transition to occur
while explaining all of the dark matter relic density.
We identify a subset of parameter space that is promising for future experiments, including the di-Higgs and Higgs signal strength
measurements at the HL-LHC and the dark matter direct detection in the XENONnT project.
}


\arxivnumber{}

\begin{document}
\maketitle

\section{Introduction}

Although the Standard Model (SM) of particle physics has been completed since the discovery of the Higgs boson at the LHC in 2012 ~\cite{ATLAS2012PLB, CMS2012PLB},
it is widely believed that new physics is required to explain various phenomena beyond the SM, including the existence of dark matter (DM) and the origin of matter-antimatter asymmetry in the Universe \cite{Planck2016}.
The most popular class of DM candidates is that of weakly interacting massive particles (WIMPs) \cite{Lee1977, Hut1977}.
These particles decoupled from the thermal bath as the early Universe was expanding and cooling, finally achieving the appropriate relic density.
In this scenario, the observed DM abundance is determined by the DM annihilation cross section, provided that the DM
particles are massive enough to become non-relativistic at freeze-out.
On the other hand, one of the widely accepted mechanisms to generate matter-antimatter asymmetry in the Universe is electroweak (EW)
baryogenesis~\cite{Kuzmin1985PLB, Cohen1990PLB, Cline2000, Morrissey2012NJP},
in which there are sufficiently large $CP$ violation (CPV) sources and the electroweak phase transition (EWPT) is sufficiently strong so that the washout of baryon
number asymmetry through sphalerons is suppressed. However, the only CPV source in the SM is provided by the CKM phase, whose effect is too small to account for the observed
asymmetry. Moreover, the EWPT is found to be a crossover rather than a first-order phase transition~\cite{Onofrio2014PRL}.

The simplest extension of the SM is to add a gauge singlet scalar. On the theoretical side, the inclusion of a singlet can
provide additional sources of CP violation, and lead to a strong first-order EWPT. 
The singlet extension of SM and the related DM and/or EWPT phenomenology have been widely studied in the literature 
\cite{Barger2009PRD,He2009PRD,Gonderinger2012PRD,Yaguna2009JCAP,Guo2010JHEP,Profumo2010PRD,Mambrini2011PRD,Cline2013PRD,Casas2017JHEP,
GAMBIT2017EPJC,Jiang2016PRD,Chiang2018PRD,Baek2012JHEP,Fairbairn2013JHEP,Li2014JHEP,Beniwal2017JHEP,Beniwal2019JHEP,Hashino2018JHEP,
Ahriche2019PLB, Alves2019JHEP, Duch2015JHEP,He2016JHEP,Chao2017JCAP,Huang2018JHEP,Kang2018JHEP,Cheng2018PRD,Kannike2019PRD,Ghorbani2019JHEP,
Ghorbani2018JPG,Profumo2007JHEP,Kozaczuk2020PRD,Musolf2019} (see refs.~\cite{Arcadi2018EPJC,Arcadi2020PR} for a recent review).
On the experimental side, the process of the strong first-order EWPT can leave a trace of stochastic background of millihertz gravitational waves (GWs). In 2015, the GW generated by the merger of binary black holes was first observed by the Advanced LIGO
experiment~\cite{LIGO2016PRL}. In the near future, the space-based interferometers such as LISA~\cite{LISA2017}, DECIGO~\cite{Sato2017JPCS},
and BBO~\cite{Crowder2005PRD} may be available to probe GWs in the range from millihertz to decihertz, which can thus be used to test
the first-order EWPT in the singlet extension of SM. The singlet scenario may also give rise to detectable signatures at the colliders, such as deviations in the triple Higgs coupling and the Higgs signal strengths.
In a recent work~\cite{Chiang2020JHEP}, we studied the two-step phase transition $(0,~w)\to (v_0,w_0)$ (type-I EWPT) in a complex singlet extension of SM with a
$\mathbb{Z}_3$ symmetry and showed that the first-order EWPT could occur if the mixing angle satisfied $|\theta|\gtrsim 11.5^{\circ}$.
The measurements of the Higgs signal strengths at the LHC, on the other hand, have restricted the mixing angle $|\theta|\lesssim 23^{\circ}$.
Future precision Higgs measurements in collider experiments, such as the high-luminosity LHC (HL-LHC) \cite{ATLAS-HL-LHC,CMS2013}, the International Linear Collider (ILC) \cite{ILC2013,Zarnecki2020},
and the Circular Electron-Positron Collider (CEPC) \cite{Yu2020EPJC} could further probe the remaining parameter space in this model~\cite{Profumo2015PRD}.

In the present work, we will continue our study in the singlet extensions of the SM, but focus on the so-called type-II EWPT. 
In this scenario, the Universe experiences a two-step phase transition.
In the first step, the singlet scalar obtains a vacuum expectation value (VEV) while the EW vacuum remains symmetric.  As the temperature decreases to the critical
temperature $T_c$ at which there exist two degenerate vacua, the second step of phase transition takes place in which the EW gauge symmetries are 
spontaneously broken by the Higgs doublet VEV $v_0$ while the singlet VEV vanishes. In addition to providing a potential barrier between the first vacuum and the EW vacuum, another 
attractive aspect of this scenario is that it is possible to introduce a new CPV source. For example, we can implement a dimension-6 effective 
operator $\mathcal{O}_6=\frac{S^2}{\Lambda^2}\bar{Q}_{3L}\tilde{H}t_R+{\rm H.c.}$ in the real scalar case with the $\mathbb{Z}_2$ symmetry to implement the additional CPV source. This can survive low-energy CPV experiment constraints \cite{Huang2018JHEP,Cline2013JCAP} due to its effects on, e.g., the 
electric dipole moment of an electron, are highly loop suppressed by a vanishing singlet VEV in the present Universe \cite{Espinosa2012JCAP}.

A scalar DM candidate can naturally arise in the type-II EWPT for the widely studied singlet extension with a $\mathbb{Z}_2$ \cite{Huang2018JHEP,Guo2010JHEP,Cline2013JCAP,Musolf2019,Espinosa2012JCAP,Curtin2014JHEP,Vaskonen2017PRD,Chao2017JCAP}
or $U(1)$ \cite{Gross2017PRL,Hashino2018JHEP,Chao2015PRD} symmetry. If the scalar VEV vanishes, such an unbroken symmetry would make the lightest neutral scalar in the dark sector stable, so that it can be a DM candidate. 
In this paper, we shall firstly revisit the $\mathbb{Z}_2$ model and perform a scan over the relevant parameter space of physical interest. 
We show the updated constraints on this model and find that the parameter space for the type-II EWPT in the singlet extensions with a $\mathbb{Z}_2$ symmetry (or a $U(1)$ symmetry) is 
largely excluded by the constraints form DM direct detections, such as LUX~\cite{LUX2017PRL}, PandaX-II~\cite{PandaX-II2017PRL}, 
and XENON1T~\cite{XENON1T2018PRL}, which strongly disfavor WIMP models with a spin-independent (SI) DM annihilation cross section larger than 
about $10^{-45}-10^{-46}~{\rm cm^2}$.
We then propose a complex singlet extension of the SM with a $CP$ symmetry in which the real component of the complex scalar can mix with the 
SM Higgs boson while the imaginary component can be a DM candidate, as protected by the $CP$ symmetry.  We will show that the type-II EWPT can 
proceed successfully while various constraints from collider and DM detection experiments can be avoided.

This work is presented as follows. In Sec.~\ref{sec:Z2Symmetry}, we study the type-II EWPT and the DM phenomenology in the real singlet extension 
of SM with a $\mathbb{Z}_2$ symmetry. In particular, we calculate the DM-nucleon scattering cross section for DM direct detection experiments in 
this model. We then perform a scan of model parameter space based on the obtained analytic results.  We comment on the model with a $U(1)$ 
symmetry toward the end of the section.
In Sec.~\ref{sec:CPsymmetry}, we introduce the model of a complex singlet extension with a $CP$ symmetry and discuss in detail its phenomenology. 
Also, we explore the type-II EWPT in this model, as well as various DM phenomena, including the invisible Higgs decay, DM relic density, 
and indirect and direct DM detections.
Finally, we summarize our findings in Sec.~\ref{sec:summary}.

\section{The Real Singlet Scalar Extension with a $\mathbb{Z}_2$ Symmetry}
\label{sec:Z2Symmetry}

\subsection{The model}

Let's first revisit the SM extension with a real singlet scalar, denoted by $S$, which interacts with SM particles through the Higgs portal.
By imposing a $\mathbb{Z}_2$ symmetry \cite{Huang2018JHEP,Guo2010JHEP,Cline2013JCAP,Musolf2019} associated with the real scalar, $S \to -S$, the most general renormalizable scalar potential is given by
\begin{eqnarray}
    V(H,S)=-\mu_{h}^2|H|^2+\lambda_h|H|^4+\frac{1}{2}\mu_s^2S^2+\frac{1}{4}\lambda_sS^4+\frac{1}{2}\lambda_m|H|^2S^2,
\end{eqnarray}
where $H$ denotes the $SU(2)$ Higgs doublet, and $\lambda_m$ is the coupling between the real scalar and the Higgs and plays important roles in both
EW phase transition and DM phenomenology. We then expand the Higgs and real scalar fields around their classical backgrounds as
\begin{equation}
H=\left(\begin{array}{c}{G^{+}} \\ {\frac{1}{\sqrt{2}}\left(h+i G^{0}\right)}\end{array}\right),~~S=s,
\label{eq:HS-expansion}
\end{equation}
where $G^{\pm}$ and $G^{0}$are the SM Goldstone bosons. In terms of the classical fields $h$ and $s$ we have the following tree-level scalar potential
at zero temperature:
\begin{equation}
    \label{eq:realtree}
    V_0(h, s)=-\frac{1}{2} \mu_{h}^{2} h^{2}+\frac{1}{4} \lambda_{h} h^{4}+\frac{1}{2}\mu_{s}^{2} s^{2}+\frac{1}{4} \lambda_{s} s^{4}
    +\frac{1}{4} \lambda_{m} h^{2} s^{2}.
\end{equation}
As shown in ref.~\cite{Cline2013JCAP}, the potential $V_0$ can give rise to a potential barrier at tree level in the type-II EWPT of this model.  
In the following we will neglect the one-loop Coleman-Weinberg potential, which just corrects the details of the phase transition but does not qualitatively change the general picture of this model.
Here we also briefly comment that for the type-I EWPT, the $\mathbb{Z}_2$ symmetry in this model enforces the existence of two local minima 
at $s=\pm w$, which prohibits a local minimum at $h=0$ \cite{Espinosa2012NPB}. Thus, it is impossible to have a tree-level barrier for the type-I EWPT.

In order to study the phase transition at finite temperature, we need to include the thermal contributions to the scalar potential.
The leading terms of one-loop finite-temperature corrections in the high-temperature expansion are given by
\begin{equation}
    V_{\rm T}(h,s,T) = \frac{1}{2}(c_h h^2+c_s s^2)T^2,
\end{equation}
where the parameters $c_h$ and $c_s$ in this model are
\begin{eqnarray}
    c_h &=& \frac{1}{48}(9g^2+3{g}'^2+12y_t^2+24\lambda_h+2\lambda_m),\nonumber\\
    c_s &=& \frac{1}{12}(3\lambda_s+2\lambda_m),
\end{eqnarray}
with $g$ and ${g}'$ the $SU(2)_L$ and $U(1)_Y$ gauge couplings while $y_t$ the top quark Yukawa coupling.
Consequently, the total effective potential is given by
\begin{equation}
    \label{eq:realtot}
    V_{\rm eff}(h,s,T) = V_0(h,s)+V_{\rm T}(h,s,T).
\end{equation}

\subsection{Phase transition}

Following the approach proposed in Refs. \cite{Espinosa2012NPB, Espinosa2012JCAP}, we add a constant term (with respect to the temperature) to the
potential~\eqref{eq:realtot} so that the total potential takes the form of Eq.~\eqref{eq:realtree} at the critical temperature $T_c$.
In this way, we have
\begin{equation}
    \label{eq:realtot2}
    V(h,s,T)=-\frac{1}{2} \mu_{h}^{2} h^{2}+\frac{1}{4} \lambda_{h} h^{4}+\frac{1}{2}\mu_{s}^{2} s^{2}+\frac{1}{4} \lambda_{s} s^{4}
    +\frac{1}{4} \lambda_{m} h^{2} s^{2}-\frac{1}{2}(c_h h^2+c_s s^2)(T_c^2-T^2).
\end{equation}
This method makes the analysis of the phase transition at the critical temperature much more convenient.
It is worth mentioning that this finite-temperature potential is gauge-independent~\cite{Jackiw1974PRD,Patel2011JHEP,Wainwright2011PRD,Katz2015PRD,Chiang2017PLB}.
We can absorb the last term of Eq.~\eqref{eq:realtot2} into the ones quadratic in {\it h} and {\it s} by defining the
{\it T}-dependent parameters
\begin{eqnarray}
    \label{eq:muh2T}
    \mu_h^2(T) &=& \mu_h^2-c_h(T^2-T_c^2),\\
    \label{eq:mus2T}
    \mu_s^2(T) &=& \mu_s^2+c_s(T^2-T_c^2).
\end{eqnarray}
As a result, one can relate the {\it T}-dependent parameters defined at the critical temperature to those at zero temperature.

The local minima of the scalar potential at the critical temperature require $\partial V_c/\partial h=0$ and
$\partial V_c/\partial s=0$ where $V_c\equiv V(T=T_c)$. For the type-II EWPT with $(0,~w)\to (v,~0)$
\footnote{Here the notation refers to the VEVs of the two scalar fields: $(\langle h \rangle, \langle s \rangle)$.}, we need the following solutions
\begin{equation}
    \label{eq:VEVs}
    w^2=-\frac{\mu_s^2}{\lambda_s};~~v^2=\frac{\mu_h^2}{\lambda_h},
\end{equation}
where $v\equiv v(T=T_c)$ and $w\equiv w(T=T_c)$. In order to ensure the vacuum stability, we demand
\begin{equation}
    \mu_s^2<0 ~,~ \lambda_h>0~{\rm and}~\lambda_s>0.
\end{equation}
Using Eqs.~\eqref{eq:muh2T} and~\eqref{eq:VEVs}, the value of the Higgs doublet VEV at the critical temperature $v$ is related to its counterpart
at zero temperature $v_{0}$ as follows
\begin{equation}
    \label{eq:EWvev}
    v_{0}^2=v^2+\frac{c_h}{\lambda_h}T_c^2.
\end{equation}
where $v_{0}\equiv v(T=0)=246~{\rm GeV}$. We see that $v<v_0=246$~GeV is aways established in this scenario because $c_h, \lambda_h > 0$.

The three elements of the scalar squared-mass matrix, evaluated at the EW gauge symmetry broken minimum, are given by
\begin{equation}
    \left.M_{h}^{2} \equiv \frac{\partial^{2} V_c}{\partial h \partial h}\right|_{b},~
    \left.M_{s}^{2} \equiv \frac{\partial^{2} V_c}{\partial s \partial s}\right|_{b},~{\rm and}~
    \left.M_{hs}^{2} \equiv \frac{\partial^{2} V_c}{\partial h \partial s}\right|_{b},
\end{equation}
where the subscript {\it b} denotes the EW gauge symmetry broken vacuum $(v,~0)$. At the critical temperature, there is another degenerate 
EW-symmetric vacuum $(0,~w)$, which would be labelled by the subscript {\it s} in the following discussions because only the $s$ field develops a nonzero VEV.
For the model considered here, 
\begin{eqnarray}
    \label{eq:tcmasses}
    M_{h}^2=-\mu_{h}^{2}+3 \lambda_{h} v^{2},~
    M_{s}^2=\mu_{s}^{2}+\frac{1}{2} \lambda_{m} v^{2},~{\rm and}~
    M_{hs}^2=0\,,
\end{eqnarray}
where $M_{h}^2>0$ and $M_{s}^2>0$ are required to ensure the existence of the local minimum at the EW broken phase.
Note that, with $\mu_s^2<0$, we require the condition
\begin{equation}
    \lambda_m>0
\end{equation}
and sufficiently large to guarantee the positivity of $M_{s}^2$. We also require
\begin{equation}
    {\rm Det}\mathcal{M}^2\equiv M_{h}^2M_{s}^2-M_{hs}^4>0.
\end{equation}
so that both mass eigenvalues of the scalar squared-mass matrix are real and positive.
Similarly, $M_{h,s}^2\equiv \partial^{2} V_c/\partial h^2|_s>0$ and $M_{s,s}^2\equiv \partial^{2} V_c/\partial s^2|_s>0$
are required to produce the local minimum at the EW symmetric vacuum.
In order for the vacuum $(0,~w)$ to be the deepest minimum along the $s$-axis,  we further impose the following condition~\cite{Espinosa2012NPB}:
\begin{equation}
    {\rm Det}\mathcal{M}^2>\frac{v^2}{w^2}M_{h}^2M_{h,s}^2.
\end{equation}
Finally, using Eqs.~\eqref{eq:muh2T},~\eqref{eq:mus2T}, and~\eqref{eq:VEVs}, we obtain the {\it T}-dependent masses
\begin{eqnarray}
    M_{h}^2(T) &=& 2\lambda_h \left [ v^2-\frac{c_h}{\lambda_h}(T^2-T_c^2) \right ],\\
    M_{s}^2(T) &=& M_{s}^{2}-\left(\frac{\lambda_{m}}{2 \lambda_{h}} c_{h}-c_{s}\right)\left(T^{2}-T_c^{2}\right).
\end{eqnarray}
The singlet mass at zero temperature is determined to be
\begin{equation}
    \label{eq:t0mass1}
    m_s^2\equiv M_s^2(T=0)=M_s^2+\left(\frac{\lambda_{m}}{2 \lambda_{h}} c_{h}-c_{s}\right)T_c^{2}.
\end{equation}

The degeneracy of the two vacua at the critical temperature implies $V_c(0,~w)=V_c(v,~0)$.
With the relations \eqref{eq:VEVs} and \eqref{eq:tcmasses} and by noting $\mu_s^2=M_s^2-\lambda_mv^2/2<0$, we obtain
\begin{equation}
    M_s^2=\frac{1}{2}v^2(\lambda_m-\sqrt{\lambda_h\lambda_s})
\end{equation}
By combining Eqs.~\eqref{eq:EWvev} and~\eqref{eq:t0mass1}, the singlet mass at zero temperature is given by
\begin{equation}
    m_{s}^{2}=\frac{1}{2}\left(\lambda_{m}-2 \sqrt{\lambda_{h} \lambda_{s}}\right) v_{0}^{2}+\left(c_{h}
    \sqrt{\frac{\lambda_{s}}{\lambda_{h}}}-c_{s}\right) T_{c}^{2}.
\end{equation}
As we shall show below, this relation requires $m_{s}$ to be smaller than about $400$~GeV.

In order to have a correct direction of the EWPT, the EW broken minimum should decrease faster than the symmetric one as the temperature drops.  
This can be expressed as the following condition
\begin{equation}
    \label{eq:ewptcond}
    \left.\frac{d \Delta V_{bs}(T)}{d T^{2}}\right|_{T_{c}}>0,
\end{equation}
where $\Delta V_{bs}(T)=V(h,s,T)|_b-V(h,s,T)|_s$. This condition can be transformed into the form
\begin{equation}
    \frac{c_{h}}{c_{s}}>\sqrt{\frac{\lambda_{h}}{\lambda_{s}}}
\end{equation}
in this model.
One can easily confirm that this condition also ensures the EW broken minimum to be the global minimum at zero temperature. 
The above general considerations of the parameters are sufficient to ensure a successful type-II EWPT in this scalar extension of the SM model.

Finally, we require the perturbative unitarity conditions \cite{Kanemura2016NPB}
\begin{equation}
    \label{eq:pertu}
    \lambda_{h}<4 \pi,~ \lambda_{s}<4 \pi,~|\lambda_{m} |<8 \pi,~
    3 \lambda_{h}+2 \lambda_{s}+\sqrt{\left(3 \lambda_{h}-2 \lambda_{s}\right)^{2}+2 \lambda_{m}^{2}}<8 \pi\,.
\end{equation}
By considering all relations discussed above, we are left with only three free parameters in this real scalar model. Here we take $\lambda_s,~\lambda_m,~{\rm and}~v/T_c$ as our input parameters and make a random scan of the following parameter space:
\begin{equation}
    10^{-3}<\lambda_s<5,~~10^{-3}<\lambda_m<5,~~{\rm and}~~1<\frac{v}{T_c}<10.
\end{equation}
Note that too large values of $\lambda_s$ and $\lambda_m$ would violate the perturbative unitarity, and the upper limit on $v/T_c$ is to guarantee the validity of high temperature expansions. Furthermore, we impose the condition $v/T_c>1$ at the critical temperature in 
order to ensure a sufficiently strong first-order EWPT~\cite{Cline2000}. We would like to mention that this condition is necessary to generate a large enough baryon asymmetry through the EW baryogenesis mechanism, since it helps suppress the washout effects caused by the sphaleron process. Some discussions concerning the reliability of this criterion can be found in Ref.~\cite{Patel2011JHEP}.

\begin{figure}
    \centering
    \includegraphics[width=75mm,angle=0]{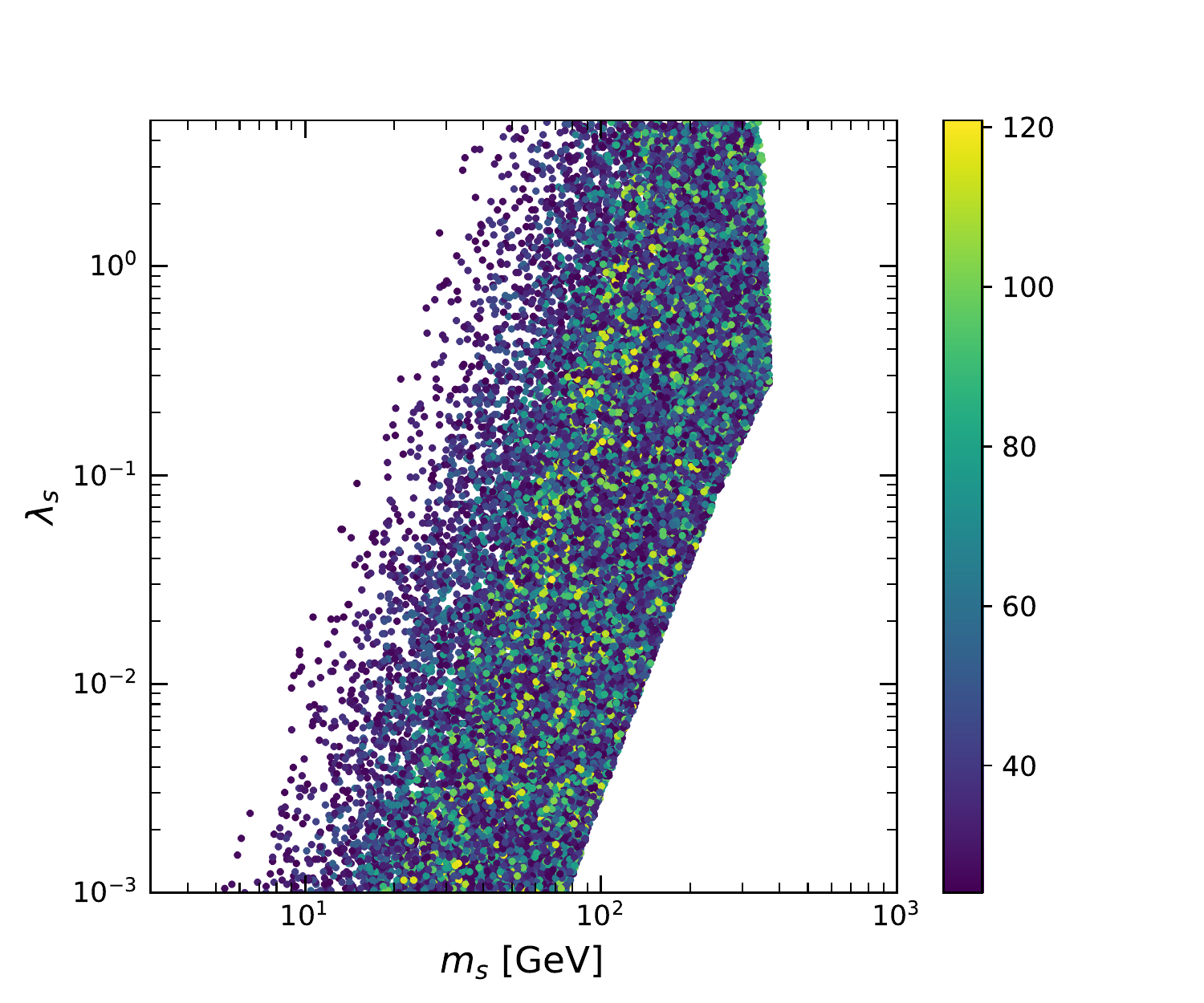}
    \includegraphics[width=75mm,angle=0]{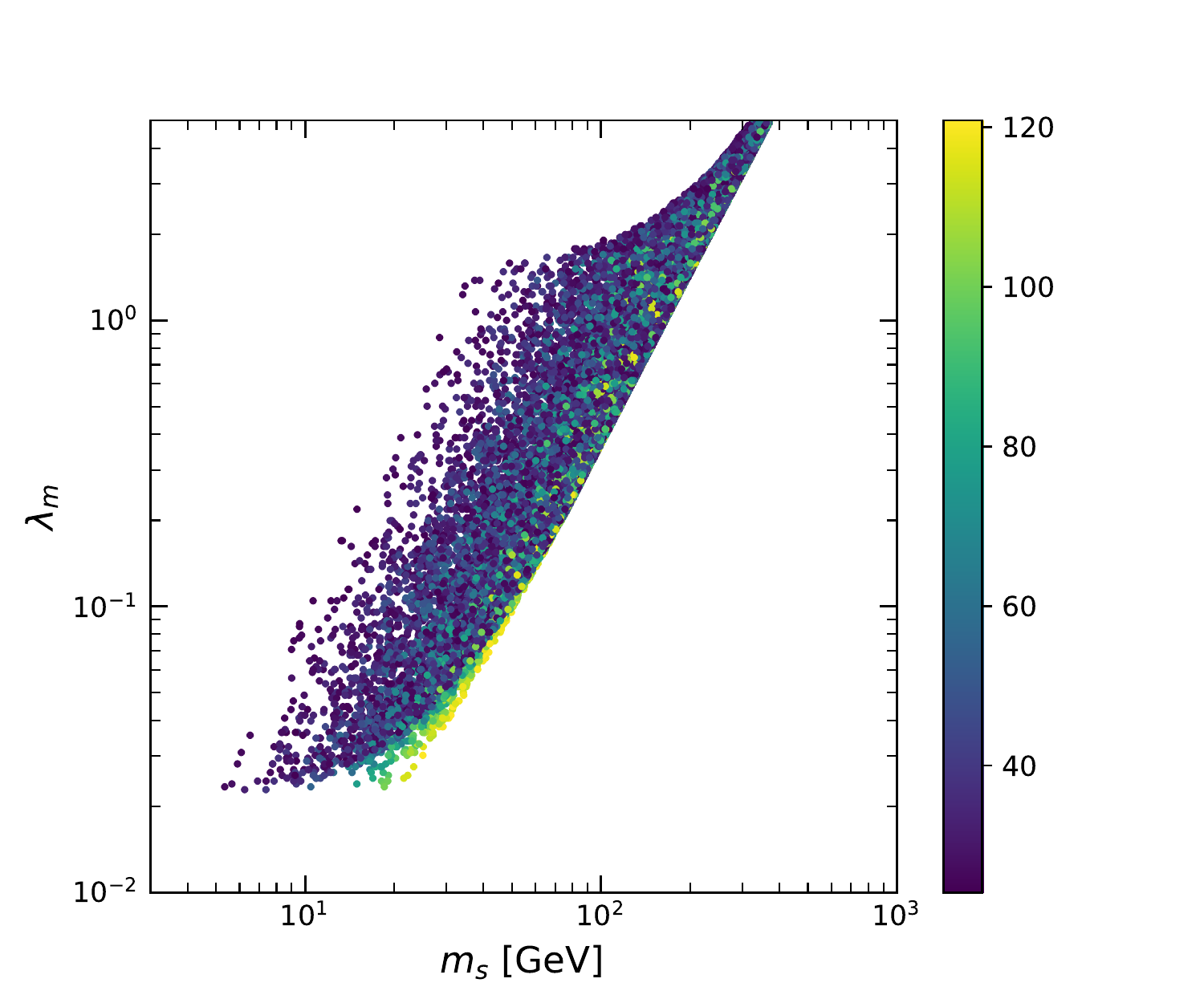}
    \caption{Scatter plots of parameters in the $\lambda_s$-$m_s$ plane (left) and the $\lambda_m$-$m_s$ plane (right) that can generate a successful type-II EWPT.  The colored bar indicates the corresponding critical temperature.}
    \label{fig:z2params}
\end{figure}

In Fig.~\ref{fig:z2params}, the colored region shows the distributions of points in the $\lambda_s-m_s$ plane (left plot) and the $\lambda_m-m_s$ plane (right plot) that can trigger a sufficiently strong first-order type-II EWPT in the model, with the color indicating the corresponding critical temperature for the chosen parameters.
The plot shows that the real scalar mass $m_s$ is bounded to be less than $\sim 400$~GeV, as alluded to before.  Furthermore, the Higgs portal coupling $\lambda_m$ should be larger than $\sim 2\times 10^{-2}$. However, such large values of $\lambda_m$ would give rise to DM-nucleon scatterings and
DM annihilation signals. Thus, as discussed below, this model suffers from strong constraints from DM detection experiments.

\subsection{Dark matter phenomenology}

After the EWPT, the VEV of the real singlet vanishes and the unbroken $\mathbb{Z}_2$ symmetry protects it as a DM candidate. 
In the standard freeze-out scenario, the DM particles are in chemical equilibrium with the other SM particles via the
annihilation-production reactions in the early Universe.
With the adiabatic expansion of the Universe, the DM population becomes nonrelativistic
and begins to decouple from the thermal bath at the time with $m_{\rm DM}/T\sim 20$ when the annihilation rate falls behind the cosmological expansion. The evolution of the DM number density is then described by the following Boltzmann equation \cite{Gondolo1991NPB}:
\begin{equation}
  \frac{d Y}{d T}=\sqrt{\frac{\pi g_{*}(T)}{45}} M_{\rm pl}\left\langle\sigma v_{\mathrm{rel}}\right\rangle
  \left[Y(T)^{2}-Y_{\mathrm{eq}}(T)^{2}\right],
\end{equation}
where the abundance $Y(T)$ ($Y_{\rm eq}(T)$) denotes the ratio of the actual (thermal equilibrium) DM number density to the entropy density,
$M_{\rm{pl}}=1.22 \times 10^{19}~\mathrm{GeV}$ is the Planck mass, $g_{\ast}$ is the effective number of
relativistic degrees of freedom, and $\left\langle\sigma v_{\mathrm{rel}}\right\rangle$ is the thermally averaged
annihilation cross section times the relative velocity. The resulting DM relic density is usually parametrized by
\begin{equation}
  h^{2}\Omega_{\rm DM}=2.742 \times 10^{8}Y_0 \frac{m_{\chi}}{\mathrm{GeV}},
\end{equation}
where $Y_0$ is the abundance of DM in the present Universe and $m_\chi$ denotes the DM mass.
The analysis of the Planck satellite's observations of the CMB radiation~\cite{Planck2016} gives $h^{2}\Omega_{\mathrm{DM}}^{\rm obs}=0.12$.
In our numerical studies, we make use of the $\textsf{MicrOMEGAs 5.0.4}$ package~\cite{Barducci2018CPC} to calculate the DM relic density.

\begin{figure}
    \centering
    \includegraphics[width=75mm,angle=0]{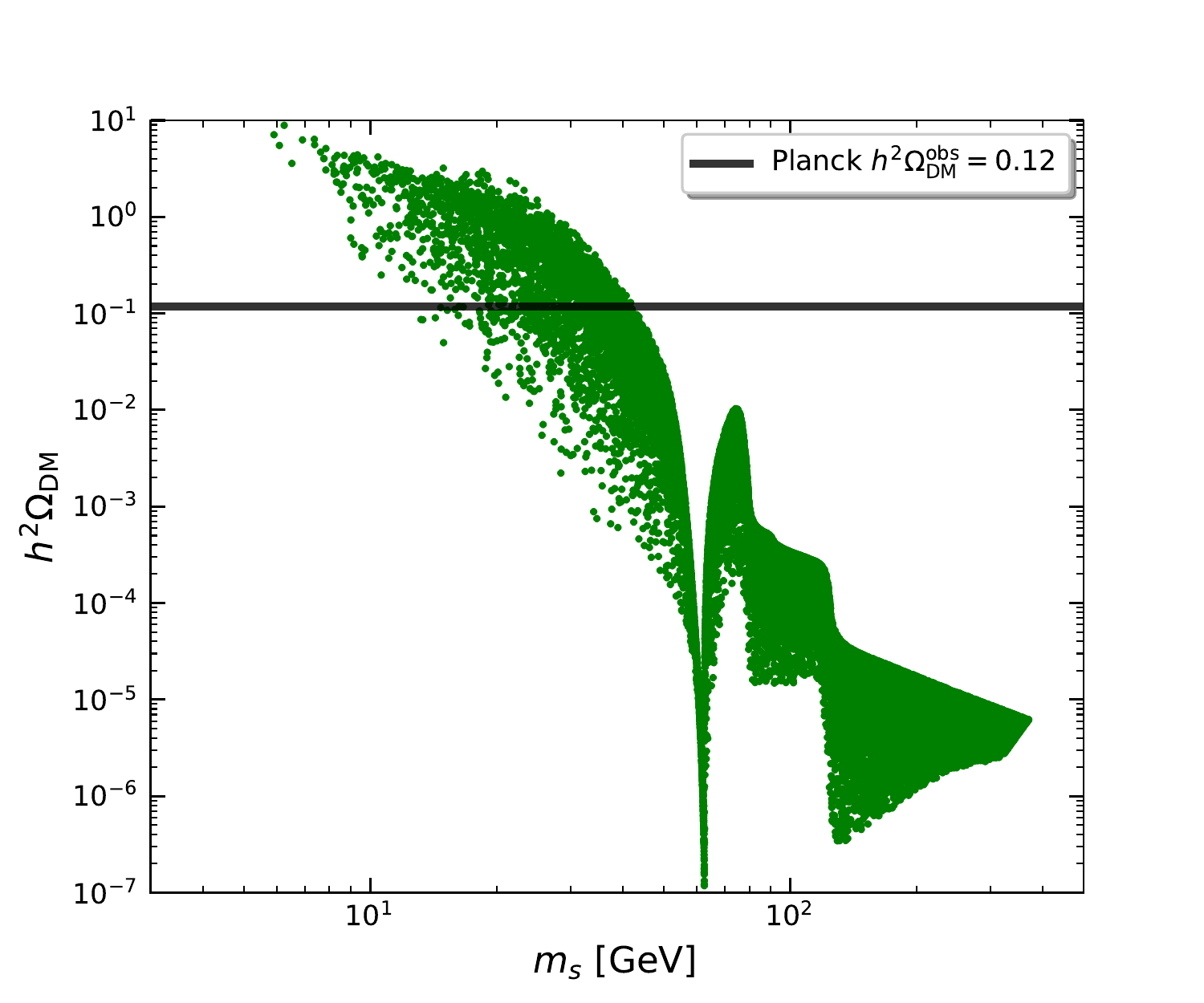}
    \includegraphics[width=75mm,angle=0]{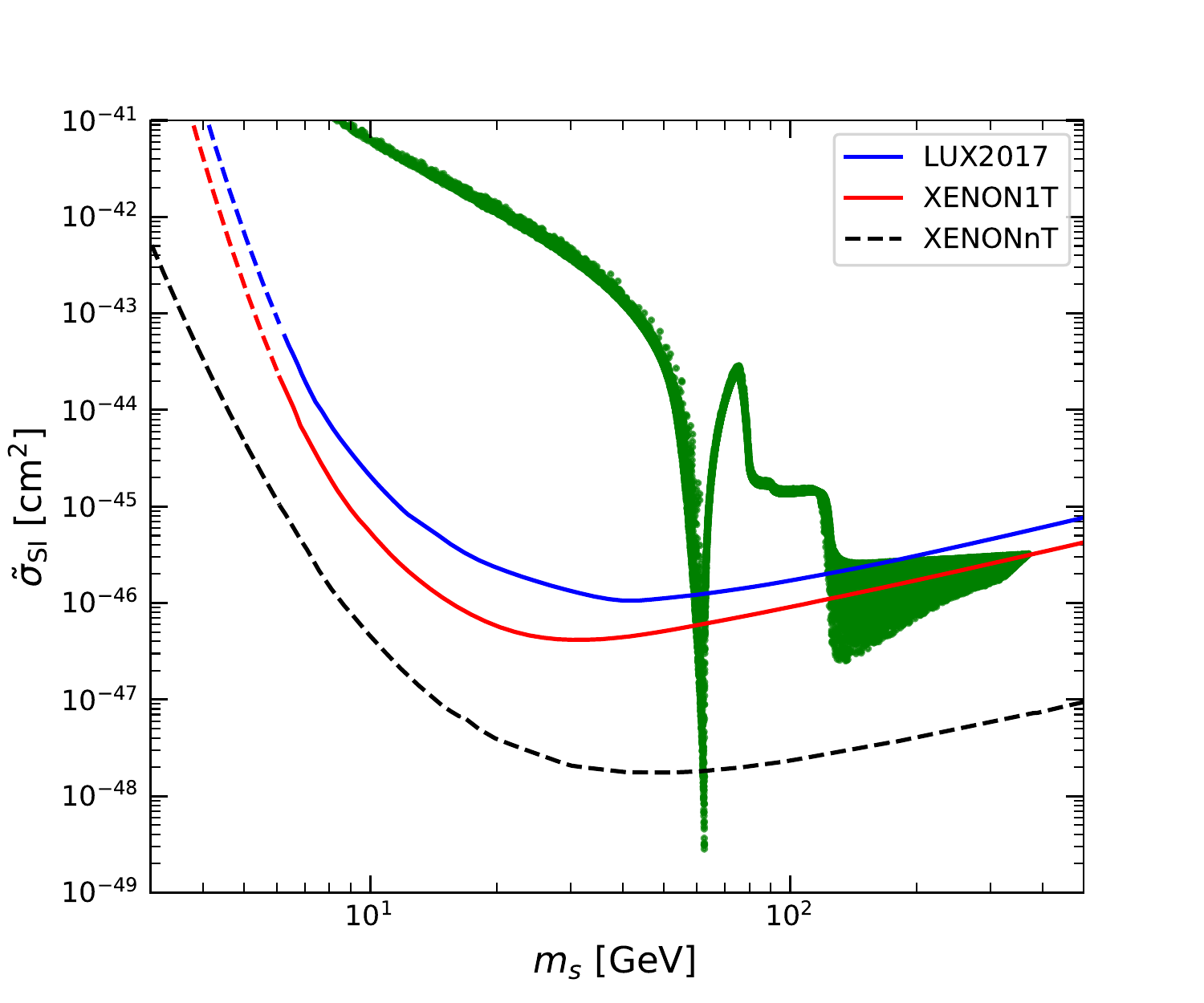}
    \caption{Left: DM thermal relic density as a function of the DM mass. The black line denotes the DM relic density derived from the Planck satellite's observationCMB radiation \cite{Planck2016}. Right: effective SI DM-nucleon elastic scattering cross section as a function of the DM mass. The blue and red curves denote the upper limits from LUX~\cite{LUX2017PRL} and XENON1T~\cite{XENON1T2018PRL} experiments, respectively. The black dashed curve represents the potential constraints from the future XENONnT project~\cite{XENONnT2016JCAP}.
    In both plots, the green scatter points represent the samples that can produce a successful type-II EWPT.}
    \label{fig:z2DM}
\end{figure}

In the left plot of Fig.~\ref{fig:z2DM}, we show the obtained DM relic density of the sample points that can trigger a successful type-II EWPT.
For $m_s\lesssim  m_{h}/2$ where $m_{h}=125$~GeV is the mass of the observed Higgs boson, a good portion of the parameter points are excluded due to their 
extremely large DM relic density, indicating that the corresponding DM annihilation rates are too small. The resonant DM annihilation 
occurs at $m_s\sim m_{h}/2$, which results in a sharp decrease of the DM relic abundance, as shown by the dip in the plot. 
When $m_s\gtrsim m_{h}$, the real scalar DM has a negligible relic density with $h^{2}\Omega_{\mathrm{DM}} \sim 10^{-6}-10^{-5}$, 
which can be explained by two reasons. For one thing, as the DM mass increases, the DM annihilation channels to the massive gauge bosons 
$W^{\pm}$ and $Z$, as well as the Higgs boson $h$ open up, leading to a decrease in the DM relic density.
The other reason lies in the fact that as $m_s$ increases a larger Higgs portal coupling
is required in order to induce the type-II EWPT, as clearly shown in the right plot of Fig.~\ref{fig:z2params}. In particular, 
when $m_s\gtrsim m_{h}$ and $\lambda_m\gtrsim 1.0$, it would lead to an efficient DM annihilation to reduce the DM density.

We now turn to the DM direct detection constraint.  For the present Higgs-portal DM models, the cross section of the SI DM-nucleon elastic scattering is given by
\begin{equation}
    \sigma_{\rm{SI}}=\frac{\lambda_m^{2} m_{N}^{4}f^{2}}{\pi m_{s}^{2} m_{h}^{4}},
\end{equation}
where $m_N$ is the nucleon mass and $f\simeq 0.3$ is the form factor \cite{Alarcon2012PRD,Cheng:2012qr,Cline2013PRD,Alarcon2014PLB,Ren2012JHEP}. 
Here we have used the $\textsf{MicrOMEGAs}$ package~\cite{Barducci2018CPC} to compute $\sigma_{\rm SI}$.
Furthermore, in order to compare with the experimental upper bounds, we have to scale the obtained scattering cross section as \cite{Cline2013JCAP}
\begin{equation}
  \label{eq:effsc}
  \tilde{\sigma}_{\rm SI}=f_{X}\sigma_{\rm SI},
\end{equation}
where the dark matter fraction
\begin{equation}
    \label{eq:fracX}
    f_{X} \equiv \frac{h^2\Omega_{\rm DM}}{h^2\Omega_{\rm DM}^{\rm obs}}.
\end{equation}
Theoretically, the dark matter fraction should be $f_{X}\leq 1$; otherwise, the Universe would be over-closed.
In our numerical calculations, however, we do not impose this restriction in the right plot of Fig.~\ref{fig:z2DM} and always apply the definition in Eq.~\eqref{eq:fracX},
since it does not affect our final numerical results by simultaneously considering the constraints from the DM relic density and DM direct detections.

The prediction of the DM-nucleon scattering cross sections for the parameter points surviving from the requirement of the type-II EWPT are displayed in the right plot of Fig.~\ref{fig:z2DM}. In the same plot, we also show the most stringent constraints from the DM direct detection experiments up to date, with the blue curve denoting the latest upper limit from the LUX experiment~\cite{LUX2017PRL} and
the red curve from the XENON1T experiment~\cite{XENON1T2018PRL}. 
It is seen that there are two relevant parameter regions which are still compatible with the strong constraints from the DM direct detections: 
one is the Higgs resonant region and the other is the region with a higher DM mass $100~{\rm GeV} \lesssim m_s \lesssim 400~{\rm GeV}$. 
Note that both allowed regions predict that the real singlet scalar has a negligible contribution to the DM relic density with a fraction of $f_{X} < 10^{-4}$, which suppresses the DM-nucleon signal to a level beyond the probe of current direct detection experiments.
Moreover, as evident in the plot, most parameter points in these two regions could be further probed or excluded by the future direct detection experiment XENONnT~\cite{XENONnT2016JCAP}, an upgraded version of XENON1T, as indicated by the black dashed curve.

Before closing this section, we note that the complex singlet scalar extension of SM with a global or local $U(1)$ symmetry (in which case the scalar 
carries a $U(1)$ charge of $q=1$) can have a potential structure much like the real singlet model with a $\mathbb{Z}_2$ symmetry.
After the type-II EWPT, this $U(1)$ symmetry would be recovered, thus protecting the complex scalar from decaying. Therefore, the complex scalar 
can be a DM candidate. In this case, the only way for the DM particle to communicate with the SM particle is via the Higgs portal coupling, 
which would also suffer from the strong constraints of DM experiments. Therefore, we expect that the final parameter space allowed by the constraints from 
the EWPT and the DM phenomenology would be similar to the real singlet case. We have confirmed this expectation with almost 
the same calculations as those performed in this section.

\section{The Complex Singlet Scalar Extension with a {\it CP} Symmetry}
\label{sec:CPsymmetry}

\subsection{The model}

From the above discussions we know that in the singlet extension with a $\mathbb{Z}_2$ symmetry or a $U(1)$ symmetry,
the type-II EWPT favors larger values of the Higgs portal coupling $\lambda_m$, which, however, suffers from the strong constraints of the DM relic density and direct detections.
In order to get out of this dilemma, we need a scalar potential with a different structure.
For this purpose, we consider the complex singlet extension of the SM with a ``{\it CP} symmetry'' $S\to S^{*}$, under which the most general renormalizable scalar potential can be written as
\begin{eqnarray}
    \label{eq:cppoten}
    V(H,S)=&-&\mu_{h}^2|H|^2+\lambda_h|H|^4-\mu_1^2(S^*S)-\frac{1}{2}\mu_2^2(S^2+S^{*2})+\lambda_1(S^*S)^2+\frac{1}{4}\lambda_2(S^2+S^{*2})^2\nonumber\\
    &+&\frac{1}{2}\lambda_3(S^*S)(S^2+S^{*2})+\kappa_1|H|^2(S^*S)+\frac{1}{2}\kappa_2|H|^2(S^2+S^{*2})+\frac{1}{\sqrt{2}}a_1^3(S+S^{*})\nonumber\\
    &+&\frac{1}{2\sqrt{2}}b_m|H|^2(S+S^{*})+\frac{\sqrt{2}}{3}c_1(S^*S)(S+S^*)+\frac{\sqrt{2}}{3}c_2(S^3+S^{*3}),
\end{eqnarray}
where all the parameters in the potential are assumed to be real. Note that the operator $(S^3+S^{*3})(S+S^*)$ has already been contained in the sixth and seventh term of Eq.~\eqref{eq:cppoten}.
By expanding the complex scalar $S$ in terms of the background fields, $S=(s+i\chi)/\sqrt{2}$, and the doublet field as in Eq.~\eqref{eq:HS-expansion}, the scalar potential can be rewritten as
\begin{eqnarray}
    \label{eq:cppoten2}
    V(h,s,\chi)=&-&\frac{1}{2}\mu_{h}^2h^2-\frac{1}{2}\mu_s^2s^2-\frac{1}{2}\mu_{\chi}^2\chi^2+\frac{1}{4}\lambda_hh^4+\frac{1}{4}\lambda_ss^4
    +\frac{1}{4}\lambda_{\chi}\chi^4+\frac{1}{2}\lambda_as^2\chi^2\nonumber\\
    &+&\frac{1}{4}\kappa_sh^2s^2+\frac{1}{4}\kappa_{\chi}h^2\chi^2+a_1^3s+\frac{1}{4}b_mh^2s+\frac{1}{3}c_ss^3+\frac{1}{3}c_{\chi}s\chi^2,
\end{eqnarray}
where
\begin{eqnarray}
    \mu_1^2&=&\frac{1}{2}(\mu_s^2+\mu_{\chi}^2),~~\mu_2^2=\frac{1}{2}(\mu_s^2-\mu_{\chi}^2),\nonumber\\
    \kappa_1&=&\frac{1}{2}(\kappa_s+\kappa_{\chi}),~~\kappa_2=\frac{1}{2}(\kappa_s-\kappa_{\chi}),\nonumber\\
    c_1&=&\frac{1}{4}(3c_s+c_{\chi}),~~~c_2=\frac{1}{4}(c_s-c_{\chi}),
\end{eqnarray}and
\begin{eqnarray}
    \lambda_1=\frac{1}{2}\left [ \frac{1}{2}(\lambda_s+\lambda_{\chi})+\lambda_a \right ],~
    \lambda_2=\frac{1}{2}\left [ \frac{1}{2}(\lambda_s+\lambda_{\chi})-\lambda_a \right ],~
    \lambda_3=\frac{1}{2}(\lambda_s-\lambda_{\chi}).
\end{eqnarray}
We assume that the pseudoscalar $\chi$ does not develop a VEV during the whole process of the type-II EWPT.
Note that we can get rid of the parameter $a_1^3$ or $c_s$ by shifting the singlet $s$ with a constant. However, we refrain from doing so for later convenience~\cite{Espinosa2012NPB}.

\subsection{Phase transition}
\label{sec:PTscan}
Following the analysis procedure in section \ref{sec:Z2Symmetry}, we rewrite the total effective potential at finite temperature as
\begin{eqnarray}
    \label{eq:cppoten3}
    V_{\rm eff}(h,s,\chi,T)=V(h,s,\chi)-\frac{1}{2}\left ( g_h h^2+g_s s^2+g_{\chi}{\chi}^2+2m_3s \right )(T_c^2-T^2),
\end{eqnarray}
where
\begin{eqnarray}
    g_h&=&\frac{3}{16}g^2+\frac{1}{16}{g}'^2+\frac{1}{4}y_t^2+\frac{1}{2}\lambda_h+\frac{1}{24}(\kappa_s+\kappa_{\chi}),\nonumber\\
    g_s&=&\frac{1}{6}\left [ \frac{1}{2}(\lambda_s+\lambda_{\chi})+\lambda_a+\kappa_s \right ]
    +\frac{1}{8}(\lambda_s-\lambda_{\chi}),\nonumber\\
    g_{\chi}&=&\frac{1}{6}\left [ \frac{1}{2}(\lambda_s+\lambda_{\chi})+\lambda_a+\kappa_{\chi} \right ]
    -\frac{1}{8}(\lambda_s-\lambda_{\chi}),\nonumber\\
    m_{3}&=&\frac{1}{12}(b_m+c_s+c_{\chi}).
\end{eqnarray}
As before, it is more convenient to combine the two parts in $V_{\rm eff}(h,s,\chi,T)$ and define four temperature-dependent parameters:
\begin{eqnarray}
    \label{eq:Tdependparams}
    \mu_h^2(T) &=& \mu_h^2-g_h(T^2-T_c^2),~~\mu_s^2(T) = \mu_s^2-g_s(T^2-T_c^2),\nonumber\\
    \mu_{\chi}^2(T) &=& \mu_{\chi}^2-g_{\chi}(T^2-T_c^2),~~a_1^3(T) = a_1^3+m_3(T^2-T_c^2).
\end{eqnarray}
With the above relations we have
\begin{eqnarray}
    \mu_h^2\equiv \mu_h^2(T_c),~\mu_s^2\equiv \mu_s^2(T_c),~\mu_{\chi}^2\equiv \mu_{\chi}^2(T_c),~{\rm and ~}a_1^3\equiv a_1^3(T_c).
\end{eqnarray}
From now on, all parameters in Eq.~\eqref{eq:cppoten2} are re-defined to be those at the critical temperature $T_c$.

The stationary points of the effective potential lie on the curves along which $\partial V_{\rm eff}/\partial \phi_i=0$
where $\phi_i=h,~s,~{\rm and}~\chi$. For the type-II EWPT, $(0,w(T),0)\to (v(T),0,0)$, we have the following conditions
\begin{equation}
    \label{eq:norel1}
    \mu_h^2(T)=\lambda_hv(T)^2,~b_m=-\frac{4a_1^3(T)}{v(T)^2},~{\rm and}~c_s=-\frac{1}{w(T)^2}(a_1^3(T)-\mu_s^2(T)w(T)+\lambda_sw(T)^3).
\end{equation}
The VEV of the EW Higgs doublet at the critical temperature $v\equiv v(T_c)$ is related to its value at zero temperature $v_0$ by
\begin{equation}
    v^2=v_0^2-\frac{g_h}{\lambda_h}T_c^2.
\end{equation}
Using the condition of vacua degeneracy at the critical temperature, i.e., $V_{\rm eff}(0,w,0,T_c)=V_{\rm eff}(v,0,0,T_c)$ where $w\equiv w(T_c)$, we derive
\begin{equation}
    \label{eq:norel2}
    a_1^3=\frac{1}{8w}\left ( \lambda_sw^4-3\lambda_hv^4+2\mu_s^2w^2\right ).
\end{equation}
The field-dependent scalar mass matrix elements are determined by ${M}_{\phi_i\phi_j}^2=\partial^2V_{\rm eff}/\partial\phi_i\partial\phi_j$. 
Due to the $CP$-odd nature of the pseudoscalar $\chi$, it cannot mix with other two scalars $h$ and $s$. On the other hand, $h$ and $s$ fields 
can mix with each other, as induced by the term proportional to $b_m$ in the potential.
Through an orthogonal rotation, we can diagonalize the mass matrix and define the following mass eigenstates: 
\begin{align}
    \begin{pmatrix}
      \mathcal{H}\\\mathcal{S}
     \end{pmatrix}=
     \begin{pmatrix}
     \cos\theta  &-\sin\theta  \\
     \sin\theta  & \cos\theta
     \end{pmatrix}\begin{pmatrix}
     h\\ s
     \end{pmatrix}.
\end{align}
At zero temperature, the mass matrix elements of $h$ and $s$ can be related to physical parameters, such as the masses of the two scalars
$m_{\mathcal{H}}=125$~GeV (the SM Higgs mass) and $m_{\mathcal{S}}$, as well as the mixing angle $\theta$, as given by the following relations:
\begin{align}
\begin{split}
  \label{eq:zeroTmass}
  M_{hh}^2&\equiv \cos^2\theta m_{\mathcal{H}}^2+\sin^2\theta m_{\mathcal{S}}^2= 2\lambda_hv_0^2,\\
  M_{ss}^2&\equiv \sin^2\theta m_{\mathcal{H}}^2+\cos^2\theta m_{\mathcal{S}}^2= -\mu_s^2(0)+\frac{1}{2}\kappa_sv_0^2,\\
  M_{hs}^2&\equiv \cos\theta \sin\theta (m_{\mathcal{S}}^2-m_{\mathcal{H}}^2)= \frac{1}{2}b_mv_0,
\end{split}
\end{align}
where $M_{\phi_i\phi_j}^2$ take values at the broken phase $(v_0,0,0)$. 
As a result of the mixing, the physical real scalar $\mathcal{S}$ is unstable and can decay into SM particles. On the other hand, the pseudoscalar $\chi$ is still stable and becomes a DM candidate since it is protected by the imposed {\it CP} symmetry.

In summary, to trigger a type-II EWPT in this complex scalar model with a {\it CP} symmetry, the model parameters should satisfy the following relations:
\begin{eqnarray}
    \label{eq:PTparams}
    \lambda_h&=&\frac{M_{hh}^2}{2v_0^2},~~ \mu_h^2=\lambda_hv^2,~~b_m=\frac{2M_{hs}^2}{v_0},~~ a_1^3=-\frac{b_mv^2}{4},\nonumber\\
    \lambda_s&=&\frac{36\lambda_hv^4+w[96a_1^3+(24M_{ss}^2+4\kappa_sT_c^2+4\lambda_aT_c^2-\lambda_{\chi}T_c^2-12\kappa_sv_0^2)w]}{12w^4-5T_c^2w^2},\nonumber\\
    \mu_s^2&=&\frac{15\lambda_hT_c^2v^4+40a_1^3T_c^2w+(24M_{ss}^2+4\kappa_sT_c^2+4\lambda_aT_c^2-\lambda_{\chi}T_c^2-12\kappa_sv_0^2)w^4}{10T_c^2w^2-24w^4},\nonumber\\
    c_s&=&-\frac{1}{w^2}(a_1^3-\mu_s^2w+\lambda_sw^3).
\end{eqnarray}
The mass of the pseudoscalar $\chi$ is given by
\begin{equation}
    \label{eq:dmmass}
    m_{\chi}^2=-\mu_{\chi}^2(0)+\frac{1}{2}\kappa_{\chi}v_0^2.
\end{equation}
From the potential given in Eq.~(\ref{eq:cppoten2}), we see that the parameters $\kappa_{\chi}$, $\lambda_a$, $c_{\chi}$ and $\lambda_{\chi}$ are directly related to the properties of the DM candidate $\chi$ so that they play important roles in the DM phenomenology. On the other hand, the same parameters would also affect significantly the type-II EWPT via their thermal contributions to the potential. Therefore, the EWPT and the DM physics are closely related to each other in the present model.

There are various theoretical constraints on the model parameters. The condition corresponding to Eq.~\eqref{eq:ewptcond} ensures a correct direction of the EWPT and requires
\begin{equation}
    g_hv^2-w(g_sw+2m_3)>0\,.
\end{equation}
Again, the same condition also warrants that the EW broken minimum at zero temperature is a global minimum.
Other bounds on the parameters have been discussed in Sec.~\ref{sec:Z2Symmetry}, a summary of which can be found in 
Table 1 of Ref.~\cite{Espinosa2012NPB}.

By taking into account the above relations and conditions, we take
\begin{equation}
    \{ w,~m_{\mathcal{S}},~\theta,~v/T_c,~\lambda_a,~\lambda_{\chi},~\kappa_a,~\kappa_{\chi},~c_{\chi},~m_{\chi} \}.
\end{equation}
as our input parameter set. 
When searching for the parameter space of the type-II EWPT, we fix
\begin{equation}
    \label{eq:fixparams1}
    \lambda_{\chi}=0.1,~~{\rm and}~~\kappa_{\chi}=0,
\end{equation}
as a concrete example and make a random scan of the other eight parameters in the following ranges:
\begin{eqnarray}\label{eq:freeparam}
    &&1<\frac{w}{\rm GeV}<2\times 10^3,~1<\frac{m_{\mathcal{S}}}{\rm GeV}<2\times 10^3,~1<\frac{m_{\chi}}{\rm GeV}<2\times 10^3,~-0.4<\theta<0.4,\nonumber\\
    &&1<\frac{v}{T_c}<10,~10^{-3}<\kappa_s<1,~10^{-3}<\lambda_a<1,~-10^{3}<\frac{c_{\chi}}{\rm GeV}<10^{3}.
\end{eqnarray}

\begin{figure}
    \centering
    \includegraphics[width=140mm,angle=0]{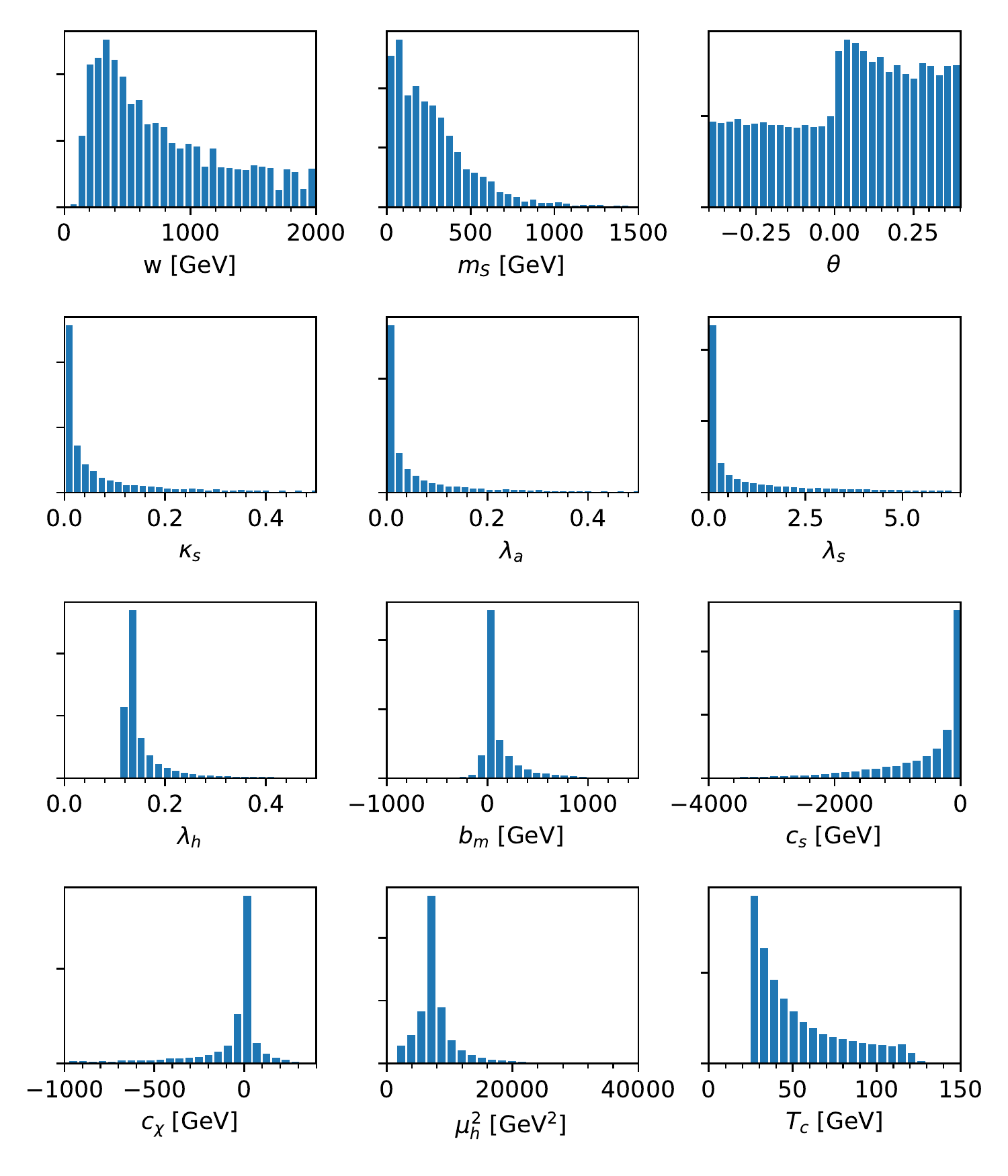}
    \caption{Distributions of parameters that can generate a sufficiently strong type-II EWPT in the $CP$-symmetric complex singlet extension of the SM.}
    \label{fig:cpdist}
\end{figure}

The model parameter space is well constrained by the current LHC measurements on various Higgs decay channels, including 
$\gamma \gamma, W W^{*}, Z Z^{*}, b \bar{b}$, and $\tau^{+} \tau^{-}$. It is remarkable~\cite{Li2014JHEP,Baek2012JHEP} that in the 
Higgs portal model all these channels have only one common Higgs signal strength:
\begin{equation}
    \mu_{\mathcal{H}}=\frac{\Gamma_{\mathcal{H}}^{\mathrm{SM}} \cos ^{4} \theta}{\Gamma_{\mathcal{H}}^{\mathrm{SM}} \cos ^{2} \theta
    +\Gamma_{\mathcal{H} \rightarrow \chi \chi}+\Gamma_{\mathcal{H} \rightarrow \mathcal{S} \mathcal{S}}}
    ~,
\end{equation}
where $\Gamma_{\mathcal{H}}^{\rm SM}$ is the total Higgs decay width in the SM and $\Gamma_{\mathcal{H} \rightarrow \chi \chi}$ and $\Gamma_{\mathcal{H} \rightarrow \mathcal{S} \mathcal{S}}$ refer to those of the exclusive channels for the SM-like Higgs decays into particles in the dark sector. It is obvious that the Higgs signal strength is suppressed by two factors: $\cos^2\theta$ and the presence of new decay channels.
In our previous work~\cite{Chiang2020JHEP}, we have shown that the Higgs signal strength measurements at the LHC restrict the mixing
angle $|\theta|\lesssim 0.4$. In this section, we have restricted the scan range of $\theta$ in accordance with this bound.

In the numerical scan, we generate one million random parameter points uniformly distributed in the range of Eq. (\ref{eq:freeparam}). 
About $\sim 3.5\%$ of the points are found to be able to trigger
a strong type-II EWPT while fulfilling other theoretical and experimental constraints. 
Furthermore, we show the distributions of the input and derived parameters relevant to the type-II EWPT in Fig.~\ref{fig:cpdist}.
The distribution of the real scalar VEV $w$ defined at the critical temperature peaks around 400~GeV and can extend up to 2~TeV.
The type-II EWPT favors a relatively light $\cal{S}$ whose mass concentrates at $m_{\cal S}\sim 200$~GeV but can be as large as $\sim 1$~TeV, in comparison with the real singlet model with a $\mathbb{Z}_2$ symmetry in Sec.~\ref{sec:Z2Symmetry}.
Also contrary to the real scalar case where the type-II phase transition requires large values of the Higgs portal coupling $\lambda_m$, most values of $\kappa_s$ in the current model are small and located in the bin of $0-0.2$.
The critical temperature $T_c$ falls in the range of $\sim 25-125$~GeV, and the triple Higgs coupling $\lambda_h$ falls mainly in the range of $0.1-0.3$.
Note that a common feature for models to generate a strong first-order EWPT is the prediction of large deviations in the cubic and quartic Higgs couplings from the SM values. 
Therefore, precision measurements of the Higgs self-couplings via di-Higgs production at
the future colliders, including HL-LHC, CEPC, and ILC, can be used to reconstruct the Higgs potential so as to confirm the nature of the 
EWPT~\cite{Alves2019JHEP,Vita2018JHEP,Adhikary2018JHEP,Maltoni2018JHEP,Borowka2019JHEP}.
Finally, the allowed ranges of some other parameters relevant to the EWPT are summarized as follows:
\begin{eqnarray}
    &&0\lesssim \lambda_a\lesssim 0.3,~0\lesssim \lambda_s\lesssim 5,~-300\lesssim \frac{b_m}{\rm GeV}\lesssim 1000,\nonumber\\
    &&-2500\lesssim \frac{c_s}{\rm GeV}\lesssim 0,~-800\lesssim \frac{c_{\chi}}{\rm GeV}\lesssim 200.
\end{eqnarray}

\subsection{Dark matter phenomenology}


In our model, the imaginary component $\chi$ of the complex scalar is found to be a DM candidate owing to the
{\it CP} symmetry of the scalar potential. From the potential in Eq.~\eqref{eq:cppoten2}, 
it is clear that there always exists a $\mathbb{Z}_2$ symmetry acting on $\chi$ as $\chi \to -\chi$, which is just another representation of the $CP$ symmetry. As noted above, the real component $s$
can mix with the Higgs boson $h$ even though it has a vanishing VEV at the final step of the type-II EWPT.
Thus, the DM candidate $\chi$ can interact with SM particles via the mediation of the mixture of both particles, leading to observable phenomena in the DM experiments.

There are thirteen parameters in the potential Eq.~\eqref{eq:cppoten2}, some of which can be related to the physical parameters $v_0$, $m_{\mathcal{H}}$, $\theta$, $m_{\mathcal{S}}$, and $m_{\chi}$ with the following relations:
\begin{eqnarray}
    \label{eq:zeroTeqs}
    \lambda_h&=&\frac{M_{hh}^2}{v_0^2},~\mu_h^2(0)=\lambda_hv_0^2,~b_m=\frac{2M_{hs}^2}{v_0},\nonumber\\
    \mu_s^2(0)&=&-M_{ss}^2+\frac{1}{2}\kappa_sv_0^2,~
    \mu_{\chi}^2(0)=-m_{\chi}^2+\frac{1}{2}\kappa_{\chi}v_0^2,\nonumber\\
    a_1^3(0)&=&-\frac{1}{4}b_mv_0^2.
\end{eqnarray}
The equations in the first and second lines of Eq.~\eqref{eq:zeroTeqs} are obtained by requiring $\partial V_{\rm eff}/\partial h=0$ in the EW symmetry broken phase $(v_0,0,0)$
of the present Universe, while the last equation of Eq.~\eqref{eq:zeroTeqs} represents the condition $\partial V_{\rm eff}/\partial s=0$ in the same phase.

we shall focus on five benchmark models with parameters summarized in Table~\ref{tab:i} in order to highlight the prominent roles played by these parameters in the dark
matter phenomenology.
In addition, other physical parameters are fixed to the following values:
\begin{equation}
    \label{eq:fixparams2}
    \lambda_{\chi}=0.1,~\kappa_{\chi}=0,~\lambda_s=0.2,~c_s=-100~{\rm GeV},~{\rm and}~\kappa_s=0.1.
\end{equation}
Note that the mixing angle $\theta$ and $c_{\chi}$ are directly related to the interaction between the DM particle $\chi$ and SM particles via $\mathcal{H}$ and $\mathcal{S}$. Hence, direct detection experiments are sensitive to these two parameters. Also, the coupling $\lambda_a$ will play a significant role in the DM annihilation when $m_{\chi}>m_{\mathcal{S}}$.
The choices of the parameters here fall in the ranges that are preferred by the type-II EWPT, as shown in Fig.~\ref{fig:cpdist}.
Our calculations below are again based upon the $\textsf{MicrOMEGAs}$ package~\cite{Barducci2018CPC}.

\begin{table}[tbp]
    \renewcommand\arraystretch{1.5}
    \centering
    \caption{\label{tab:i} A summary of parameters for the five benchmark models.}
    \begin{tabular}{|c|c|c|c|c|c|}
    \hline
    Model & $\theta$ & $\lambda_a$ & $c_{\chi}$ {\rm [GeV]} \\
    \hline
    A & 0.1 & 0.5 & $-100$ \\
    B & 0.1 & 0.1 & $-100$ \\
    C & 0.3 & 0.5 & $-100$ \\
    D & 0.1 & 0.5 & $-300$ \\
    E & 0.2 & 0.3 & $-200$ \\
    \hline
    \end{tabular}
\end{table}

\subsubsection{Invisible Higgs decay}

\begin{figure}
    \centering
    \includegraphics[width=110mm,angle=0]{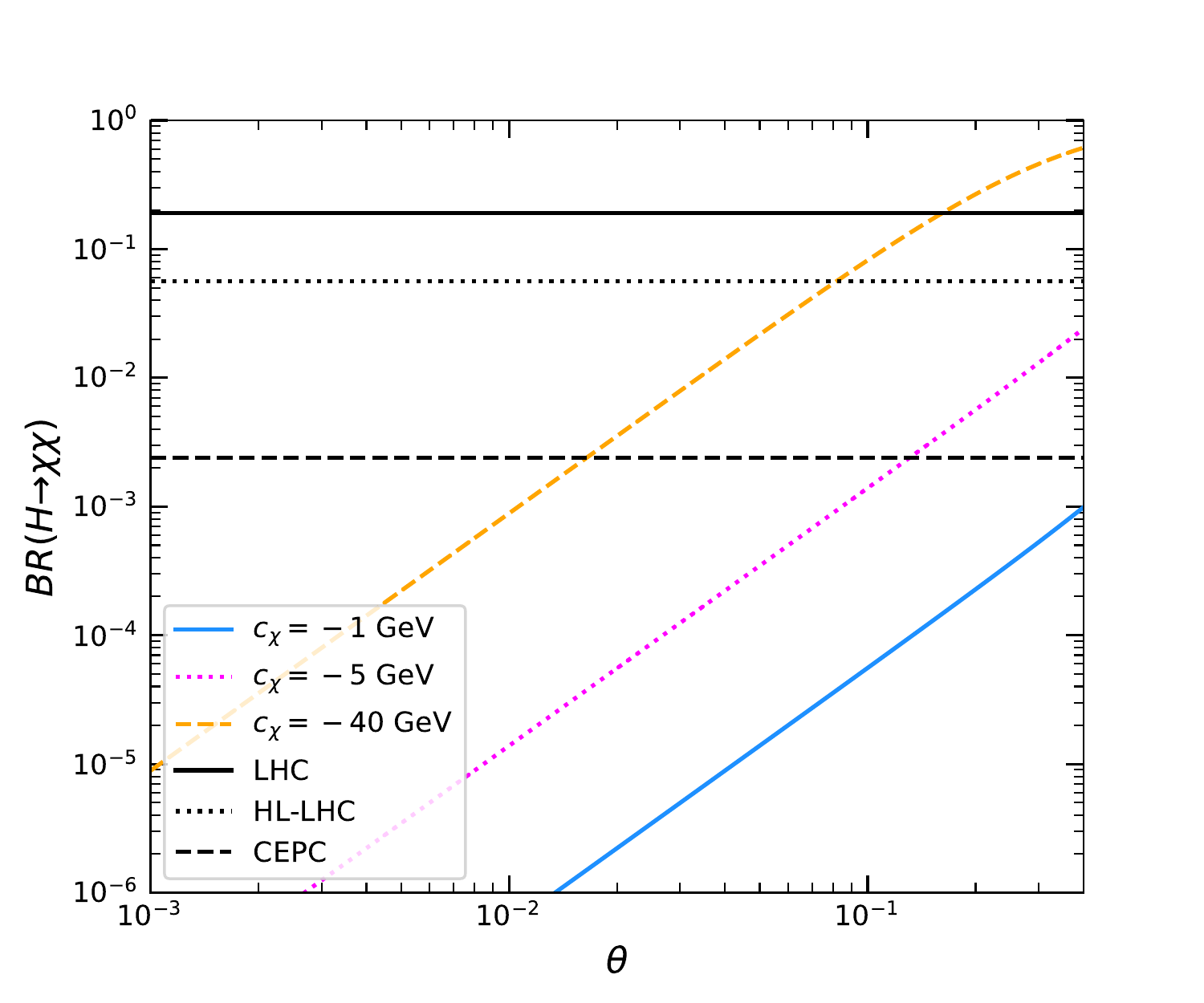}
    \caption{Invisible Higgs decay branching ratio as a function of $\theta$.
    The blue solid line, pink dotted line, and yellow dashed line correspond respectively to $c_{\chi}=-1$~GeV, $c_{\chi}=-5$~GeV, and $c_{\chi}=-40$~GeV.
   In this plot, we fix the DM mass $m_{\chi}=20$~GeV, the real scalar mass $m_{\mathcal{S}}=400$~GeV, and the coupling $\lambda_a=0.5$, with the other parameters given in Eq.~\eqref{eq:fixparams2}.
    The black solid, dotted, and dashed lines correspond respectively to the upper limit on the invisible Higgs decay branching ratio from the current LHC analysis~\cite{Sirunyan2019PLB}, the future HL-LHC~\cite{Liu2017CPC} and CEPC~\cite{Tan2020}.}
    \label{fig:decay1}
\end{figure}
For $m_{\chi}<m_{\mathcal{H}}/2$, the $\mathcal{H}\to \chi\chi$ decay is kinematically allowed.
Its partial decay width is
\begin{equation}
    \label{eq:decayhxx}
    \Gamma_{\mathcal{H}\to \chi\chi}=\frac{\lambda_{H\chi\chi}^2}{32\pi m_{\mathcal{H}}}\sqrt{1-\frac{4m_{\chi}^2}{m_{\mathcal{H}}^2}},
\end{equation}
where
\begin{equation}
    \label{eq:chxx}
    \lambda_{\mathcal{H}\chi\chi}= -\kappa_{\chi}v_0\cos\theta+\frac{2}{3}c_{\chi}\sin\theta.
\end{equation}
In Eq.~(\ref{eq:chxx}), the first term proportional to $\kappa_{\chi}$ represents the widely studied Higgs portal originated from the $h^2\chi^2$ operator in the
potential~\eqref{eq:cppoten2}, and the second term stems from the mixing between $h$ and $s$ and hence the suppression factor of mixing angle.
For simplicity, we focus on the case with $\kappa_{\chi}=0$, i.e., we assume $\kappa_1=\kappa_2$ for the initial potential in Eq.~\eqref{eq:cppoten}.
Fig.~\ref{fig:decay1} shows the $\mathcal{H}\to \chi\chi$ branching ratio as a function of $\theta$. In this plot, we have assumed the DM mass $m_{\chi}=20$~GeV, 
the real scalar mass $m_{\mathcal{S}}=400$~GeV, and the coupling $\lambda_a=0.5$.
The other parameters are given in Eq.~\eqref{eq:fixparams2}.  In the SM, the dominant Higgs decay channel is $\mathcal{H}\to b\bar{b}$.
However, when the invisible Higgs decay opens up, the ratio of the invisible decay width to the $\mathcal{H}\to b\bar{b}$ decay width is estimated to be
$R\simeq \lambda_{\mathcal{H}\chi\chi}^2/(\lambda_f^2m_{\mathcal{H}}^2)=1.7c_{\chi}^2\sin^2\theta/(m_b^2\cos^2\theta)$, where
$\lambda_f=m_b\cos\theta/v_0$ and $m_b\simeq 4$~GeV is the bottom quark mass~GeV~\cite{Tanabashi2018PRD}. Thus, for $\theta\sim 0.1$ and $|c_{\chi}|\sim 40$~GeV,
the invisible decay becomes competitive with the $b\bar{b}$ channel, which can be seen in Fig.~\ref{fig:decay1}.
Currently, the upper limit on the branching ratio of the invisible Higgs decay is 19\% by the LHC~\cite{Sirunyan2019PLB} (black solid line), and this limit can be further
improved to 5.6\% at HL-LHC~\cite{Liu2017CPC} (black dotted line) and 0.24\% at CEPC~\cite{Tan2020} (black dashed line).

\begin{figure}
    \centering
    \includegraphics[width=140mm,angle=0]{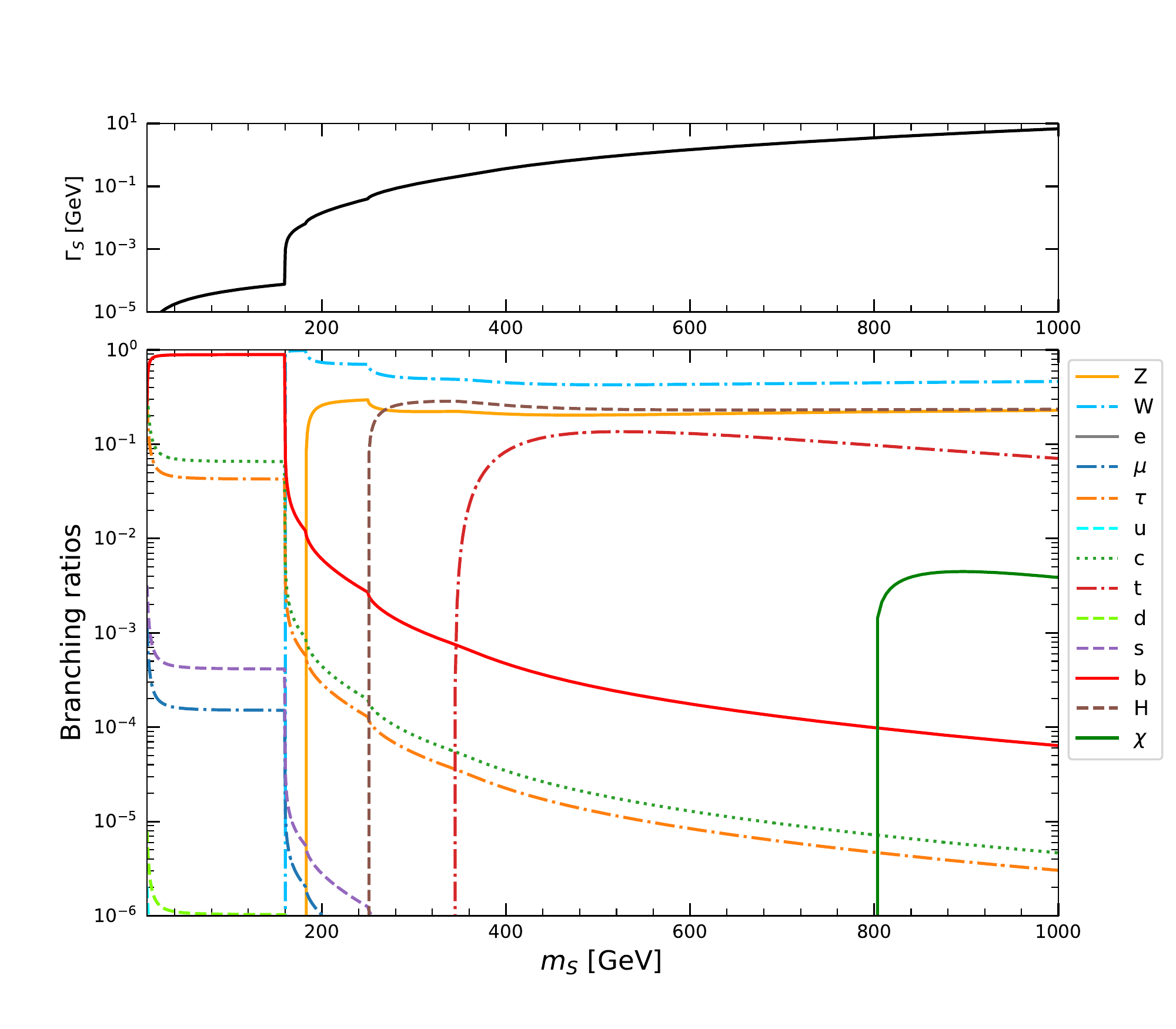}
    \caption{Branching ratios of $\mathcal{S}$ decays (lower plot) and its total decay width (upper plot) as a function of $m_{\mathcal{S}}$.
    Here DM mass $m_{\chi}=400$~GeV and the other parameters are given in Eq.~\eqref{eq:fixparams2} and model A.}
    \label{fig:decay3}
\end{figure}

When $m_{\mathcal{S}}<m_{\mathcal{H}}/2$, the $\mathcal{H}\to \mathcal{S}\mathcal{S}$ decay process can also contribute to the invisible Higgs 
decay if $\mathcal{S}\to \chi\chi$ is kinematically allowed. In this case, the total invisible Higgs decay branching ratio is given by
\begin{eqnarray}
    BR({\rm inv}) = BR(\mathcal{H}\to \chi\chi)
                   +BR(\mathcal{H}\to \mathcal{S}\mathcal{S})BR(\mathcal{S}\to \chi\chi)^2.
\end{eqnarray}
As in Eq.~\eqref{eq:decayhxx}, the partial decay width $\Gamma_{\mathcal{H}\to \mathcal{S}\mathcal{S}}$ is proportional to the square of
the coupling
\begin{eqnarray}
    \lambda_{\mathcal{H}\mathcal{S}\mathcal{S}}&=&-\kappa_sv_0\cos^3\theta-(b_m-2c_s)\cos^2\theta\sin\theta \nonumber\\
    &&+\frac{1}{2}\sin^2\theta[4(\kappa_s-3\lambda_h)v_0\cos\theta+b_m\sin\theta].
\end{eqnarray}
With $m_{\chi}=20$~GeV, $m_{\cal{S}}=50$~GeV, $c_{\chi}=-5$~GeV, and other parameters given in Eq.~\eqref{eq:fixparams2}, we find the branching ratio of invisible Higgs decays in the range $\sim 10^{-5}-10^{-3}$ for $0.01\lesssim \theta \lesssim 0.1$.

Fig.~\ref{fig:decay3} shows the total decay width of $\cal{S}$ in the upper plot, and its respective branching ratios to SM particles and to $\chi$ in the 
lower plot. Here we fix the DM mass to be $m_{\chi}=400$~GeV, and other parameters are given in Eq.~\eqref{eq:fixparams2} and model A 
in Table~\ref{tab:i}. We see that when $m_{\mathcal{S}}< 2m_{W}$, the process $\mathcal{S}\to b\bar{b}$ is the dominant
channel of $\mathcal{S}$ decays. As $m_{\mathcal{S}}$ increases, the $W^+W^-$, $ZZ$, $\cal HH$, and $t\bar{t}$ channels open
up successively and dominate over the $\mathcal{S}$ decays. When $m_{\mathcal{S}}> 2m_{\chi}$, the channel $\mathcal{S}\to \chi\chi$
becomes available. But, by comparing with $\mathcal{S}$ decays to SM particles, this process only contributes a small fraction of
$\simeq 5\times 10^{-3}$ and thus plays a subdominant role. 
For $m_{\mathcal{S}}$ around $\sim 100$~GeV,
the total decay width of $\mathcal{S}$ is estimated to be $\Gamma_{S}\simeq 6\times 10^{-5}~{\rm GeV}\simeq 10^{-2}\Gamma_{H}$, where $\Gamma_{H}$
is the total decay width of the SM Higgs boson. This is because the real scalar $\mathcal{S}$ decaying to SM particles is suppressed by a factor of
$\sin^2\theta$ in comparison with the SM Higgs decays. The $\mathcal{S}$ total decay width can go up to $\sim 10$~GeV with its mass increasing to $\sim 1$~TeV.
Note again that due to the mixing, $\mathcal{S}$ could induce observable effects sensitive to $\theta$ at the LHC. Here we have set $\theta=0.1$, 
which is allowed by the Higgs signal strength restriction
$|\theta|\lesssim 0.4$. The future hadron and lepton colliders, such as HL-LHC and ILC, could probe the mixing angle $\theta$ in the range 
of $0.1-0.4$ and provide more opportunities to test our scenarios.

\subsubsection{Relic density}

\begin{figure}
    \centering
    \includegraphics[width=110mm,angle=0]{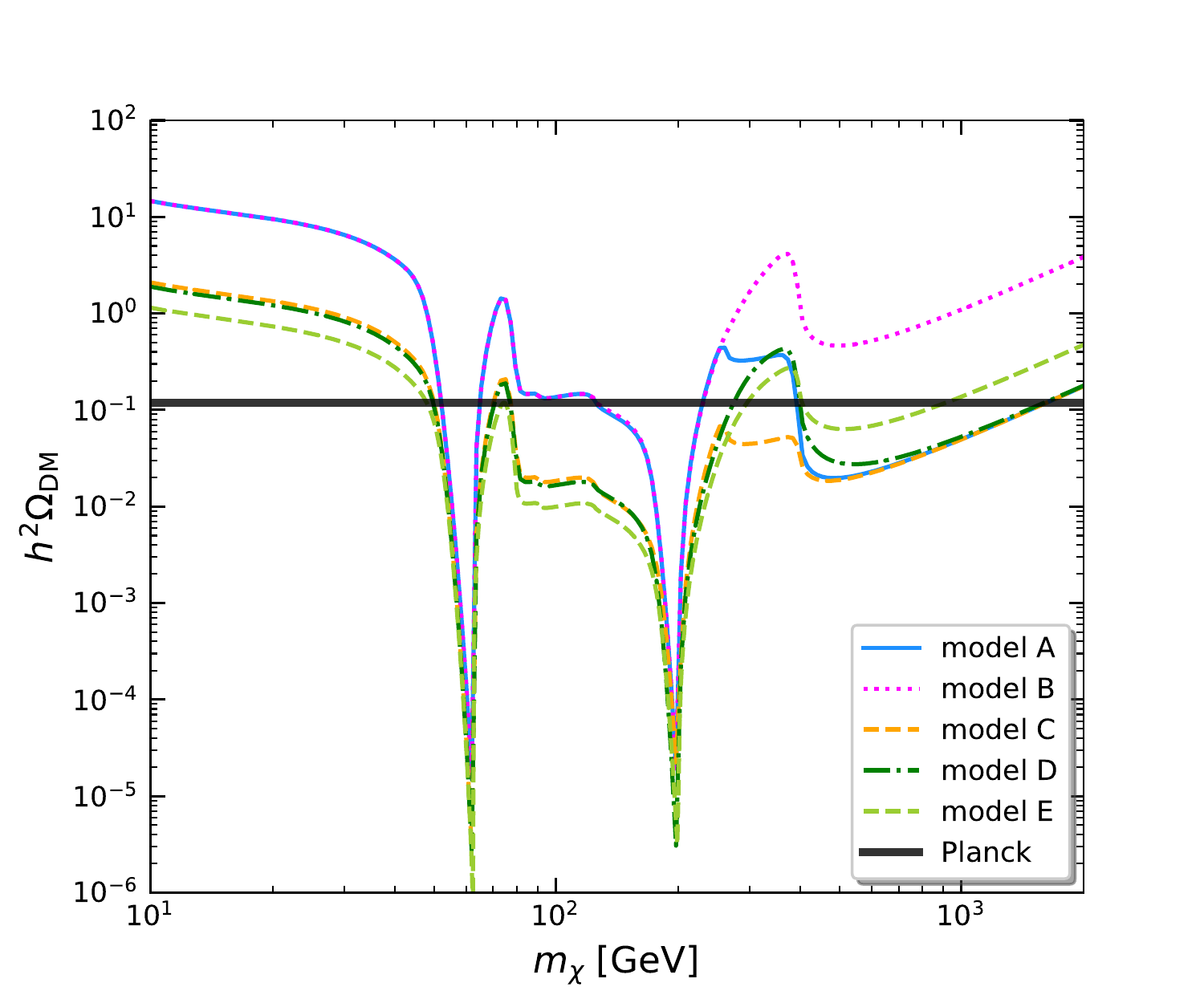}
    \caption{DM relic density in the five benchmark models as a function of $m_{\chi}$.}
    \label{fig:cprd}
\end{figure}

In Fig.~\ref{fig:cprd}, we show the DM relic density in the five benchmark models as a function of $m_{\chi}$.
For $m_{\chi}<m_{\mathcal{H}}/2$, the processes contributing to the DM annihilation are those with the SM fermions $f$ in the final states, $\chi\chi \to f\bar{f}$, mediated by $\mathcal{H}$
and $\mathcal{S}$. 
Still, the relevant parameters are $\theta$ and $c_{\chi}$ (see Eq.~\eqref{eq:chxx}) and the DM annihilation cross sections are proportional to the product $|\sin\theta\times c_{\chi}|^2$.
As $m_\chi$ increases, the DM relic density can be reduced greatly due to the allowance of annihilation into pairs of the SM weak gauge bosons and Higgs bosons, as well as the one into $\mathcal{H}\mathcal{S}$.
The process $\chi\chi\to \mathcal{H}\mathcal{S}$ is kinematically allowed when
$2m_{\chi}\gtrsim m_{\mathcal{H}}+m_{\mathcal{S}}$.
The resonant DM annihilations occur at $m_{\chi}\simeq m_{\mathcal{H}}/2$ and $m_{\mathcal{S}}/2$, thus sharp decrease of the DM relic density at the two dips in the curves.
The final decrease in the relic density takes place when $m_{\chi}>m_{\mathcal{S}}=400$~GeV, since
$\chi\chi\to \mathcal{S}\mathcal{S}$ dominates over the other annihilation processes. In this case, $\lambda_a$ controls the DM freeze-out process in the five benchmark models. 
In particular, models A, C, and D share the same large coupling $\lambda_a=0.5$, which can boost the cross section of $\chi\chi \to \mathcal{S} \mathcal{S}$ 
so that it leads to a DM relic density consistent with the current cosmological observations in the DM mass range between 400~GeV and 1.6~TeV. 
Model E with $\lambda_a=0.3$ can also generate a DM relic density $\lesssim 0.12$ in the DM mass range of $400-900$~GeV.
On the other hand, in model B, $\lambda_a$ is fixed at $0.1$ that is insufficient to suppress the relic density.

\begin{figure}
    \centering
    \includegraphics[width=140mm,angle=0]{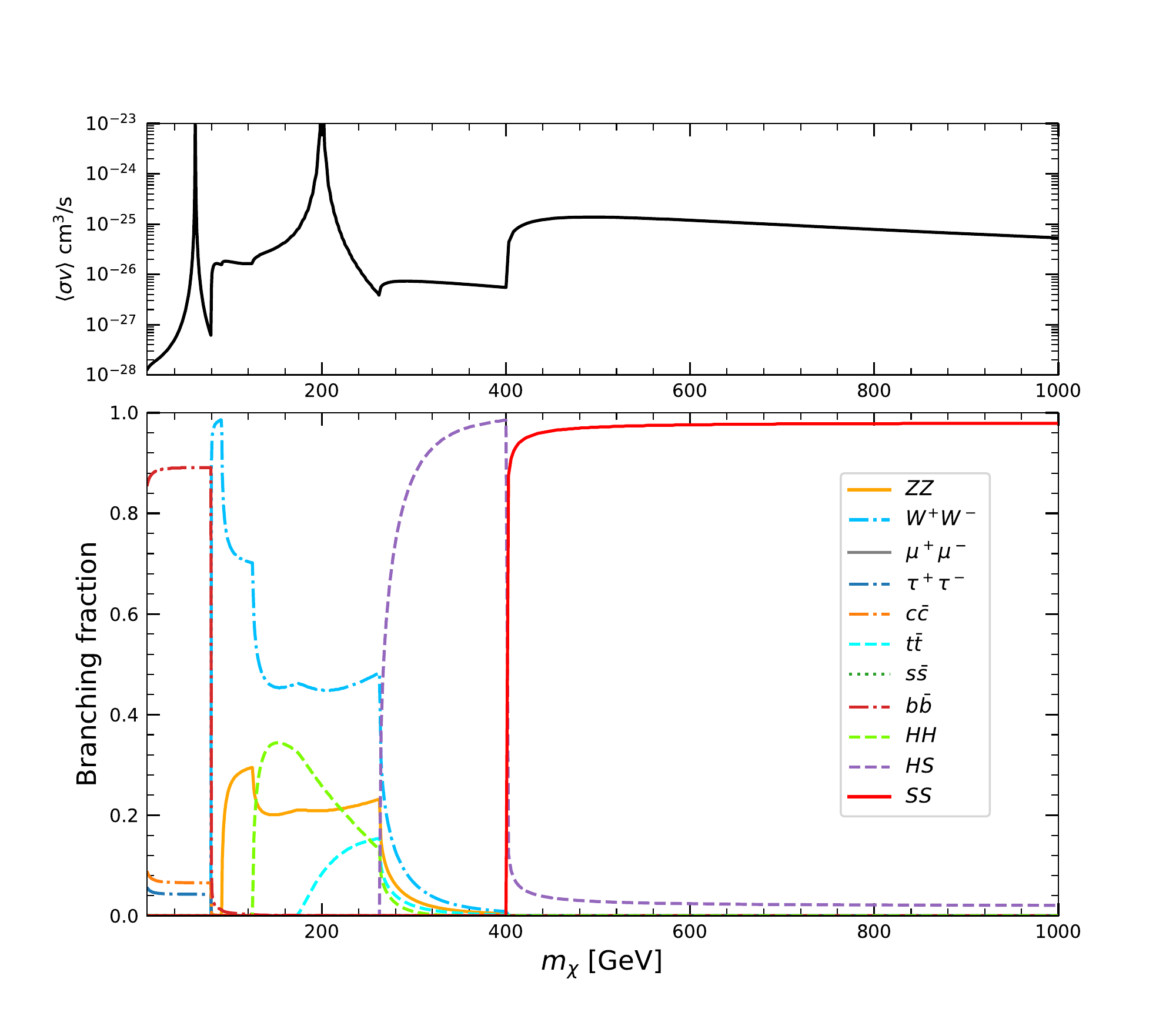}
    \caption{Upper plot: Total DM annihilation cross section as a function of DM mass in the present Universe.
    Lower plot: Branching ratios DM annihilation to various final states as a function of DM mass. The real scalar mass $m_{\mathcal{S}}=400$~GeV, and the other parameters are given in Eq.~\eqref{eq:fixparams2} and model A.}
    \label{fig:cpannBF}
\end{figure}
\begin{figure}
    \centering
    \includegraphics[width=110mm,angle=0]{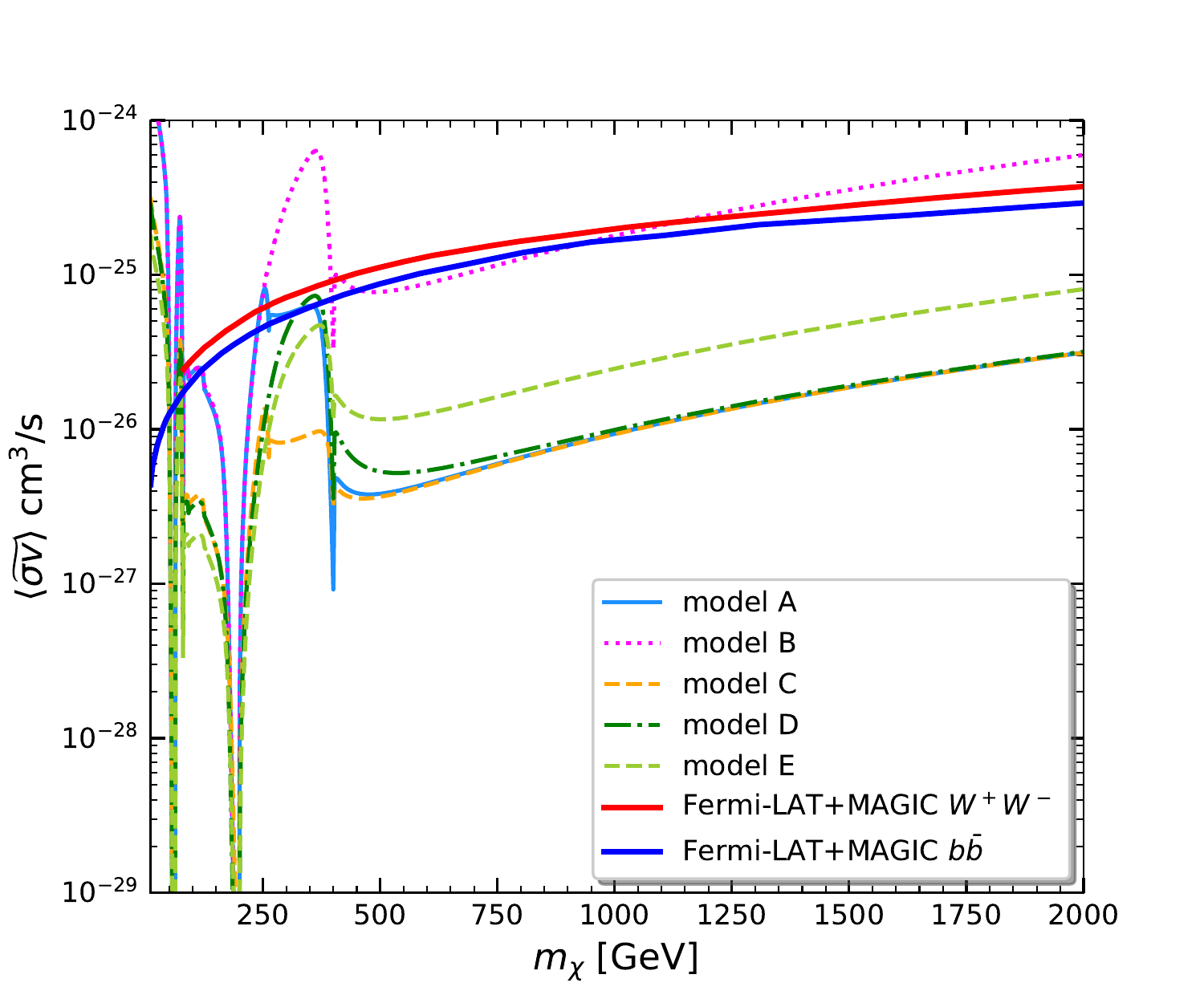}
    \caption{Effective annihilation cross section in the five benchmark models as function of DM mass.
    The red and blue curves represent respectively the constraints on the DM annihilation to $W^+W^-$ and $b\bar{b}$ final states from a combined analysis of Fermi-LAT \cite{Fermi2015PRL} and MAGIC \cite{MAGIC2016JCAP} observations of dwarf satellite galaxies.}
    \label{fig:cpsv}
\end{figure}

\subsubsection{Indirect detection}

We now turn to the study of DM indirect detections via the measurements of cosmic rays. For this purpose, in Fig.~\ref{fig:cpannBF}
we calculate the total DM annihilation cross section in the present Universe (upper plot) and its branching ratios to various final states (lower plot). Here, we take parameters in model A as an example.
In the calculation of the total annihilation cross section, the polarizations of the $W$ and $Z$ bosons and the photon radiation from 3-body final states are also considered. As we have observed in the DM relic density computations, there are two resonant annihilation regions at around $m_{\chi}=m_{\mathcal{H}}/2$ and $m_{\chi}=m_{\mathcal{S}}/2$. Moreover, the annihilation cross section becomes nearly a constant
$\approx 5\times 10^{-26}~{\rm cm^{-3}/s}$ when $m_{\chi}>m_{\mathcal{S}}=400$~GeV. From the annihilation branching ratios, we observe that for $m_{\chi}<(m_{\mathcal{H}}+m_{\mathcal{S}})/2$ the annihilation to SM final states $b\bar{b}$, $W^+W^-$, $ZZ$, $HH$, and
$t\bar{t}$ are the dominant processes. When $m_{\chi}>(m_{\mathcal{H}}+m_{\mathcal{S}})/2$, the process $\chi\chi\to \mathcal{H}\mathcal{S}$
assumes the main contribution to annihilation. Finally, the $\mathcal{S}\mathcal{S}$ final state dominates the annihilation when
$m_{\chi}>m_{\mathcal{S}}$.

In order to compare with the experimental upper limits, we define the effective annihilation cross section as follows:
\begin{equation}
    \left \langle \widetilde{\sigma v} \right \rangle =f_X^2\left \langle \sigma v \right \rangle,
\end{equation}
which is plotted in Fig.~\ref{fig:cpsv} for the five benchmark models. 
In the same plot, the solid red and blue lines represent respectively the constraints on the DM annihilations to the $W^+W^-$ and $b\bar{b}$ final states from a combined analysis of Fermi-LAT \cite{Fermi2015PRL} and MAGIC \cite{MAGIC2016JCAP} observations of dwarf satellite galaxies. It is seen that model B has been restricted severely by these
gamma-ray observations due to its relatively large value of DM relic density, while the other models can survive them.
Note that such considerations of DM indirect detection limits are conservative since the cross section of a specific DM annihilation channel is always less than the total one.
We also note that the reason why we have applied the constraints from the $b\bar{b}$ and $W^+W^-$ final states is that the former channel dominates the DM annihilation when $m_\chi < m_W$ while the latter decay mode becomes relevant in the large DM mass regime when the annihilation $\chi\chi \to \mathcal{S} \mathcal{S}$ opens up, with ${\cal S}$ mostly decaying into $W^+ W^-$ for the present parameter choice, 
as seen in Fig.~\ref{fig:decay3}. In fact, a more appropriate upper limit for the $m_\chi>m_{\cal S}$ case should be given by the DM annihilation process $\chi \chi \to 2W^+ + 2W^-$. However, we cannot find the DM indirect detection constraints on this $4W$ final state in the literature, and have thus applied the bounds on the $W^+ W^-$ channel instead. Nevertheless, one can expect that the two 
constraints can differ at most by a factor of $2$~\cite{Jin2013JCAP,Clark2018PRD}.

\subsubsection{Direct detection}

\begin{figure}
    \centering
    \includegraphics[width=110mm,angle=0]{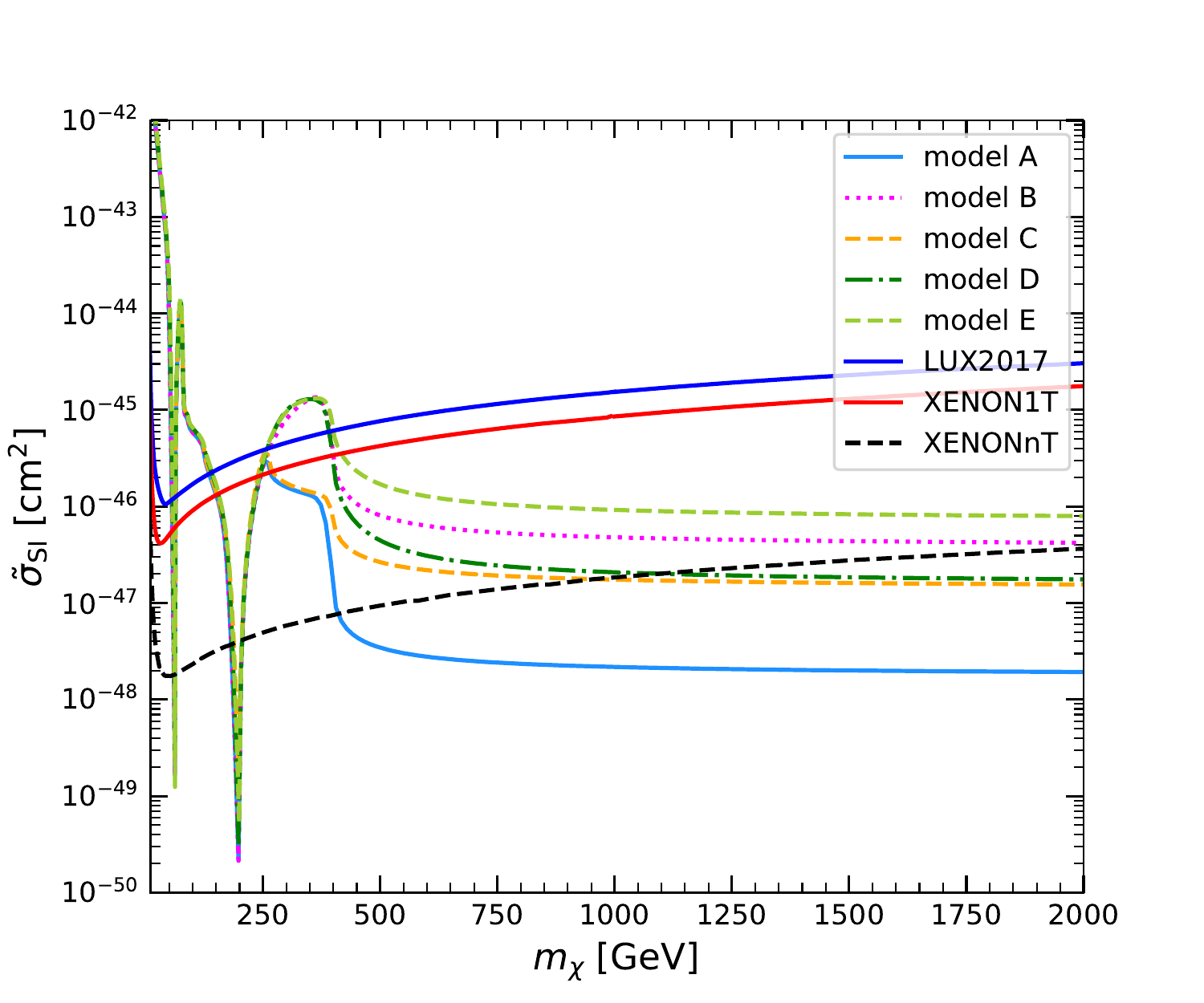}
    \caption{Effective SI DM-nucleon elastic scattering cross section in the five benchmark models as a function of the DM mass.
    The blue and red curves denote the upper limits on scattering
    cross section from LUX~\cite{LUX2017PRL} and XENON1T~\cite{XENON1T2018PRL} experiments, respectively.
    The black dashed curve represents the potential constraints from the future XENONnT project~\cite{XENONnT2016JCAP}.}
    \label{fig:cpdd}
\end{figure}

The DM direct detection tries to measure the DM-nucleon scatterings in the deep underground laboratories. 
In the present model, the DM-nucleon interaction can proceed via the $\mathcal{H}$ and $\mathcal{S}$ portals, which are, however, suppressed by the mixing angle $\theta$.
In Fig.~\ref{fig:cpdd}, we plot the rescaled cross section of DM-nucleon scatterings $\tilde{\sigma}_{\rm SI}$ defined in Eq.~\eqref{eq:effsc} as a function of DM mass. 
As shown in the figure, for $m_{\chi}\lesssim m_{\mathcal{S}}=400$~GeV, all the benchmark models are constrained by LUX and XENON1T, except for 
the DM mass sitting at the $\mathcal{H}$ and $\mathcal{S}$ resonances. As for the heavy DM regime with $m_{\chi}>m_{\mathcal{S}}$, 
the effective DM-nucleon scattering cross section of all models is suppressed to the level that can evade the current direct detection constraints. 
In order to probe the large DM mass region, it is widely believed that the near-future XENONnT project can play an important role with its 
expected sensitivity shown as the black dashed curve in Fig.~\ref{fig:cpdd}. As a result, the XENONnT experiment can test and constrain the 
entire DM mass range in the plot for models B and E, a correlated result of the large DM relic density predicted in the model. 
In contrast, the XENONnT constraint on model A is the weakest.  This can be understood as follows: the DM relic density in the large mass region 
is controlled by the coupling $\lambda_a$, while the DM direct detection signal is induced via the mediation of $\cal{H}$ and $\cal{S}$ and can 
only place limits on $|\sin\theta\times c_{\chi}|$. Therefore, models A, C, and D predict the same DM relic density due to the same value of $\lambda_a$, 
but, due to their different values of $|\sin\theta\times c_{\chi}|$, model A can totally avoid the probe of XENONnT, while models C and D can be 
tested when the DM is lighter than 1~TeV.  Interestingly, with the parameter choices in Table~\ref{tab:i}, the pseudoscalar $\chi$ with
$m_{\chi}\sim 1.6$~TeV in models A, C, and D, and $m_{\chi}\sim 0.9$~TeV in model E can constitute all the DM density in our Universe, 
while being consistent with all the current DM experimental constraints.

\begin{figure}
    \centering
    \includegraphics[width=100mm,angle=0]{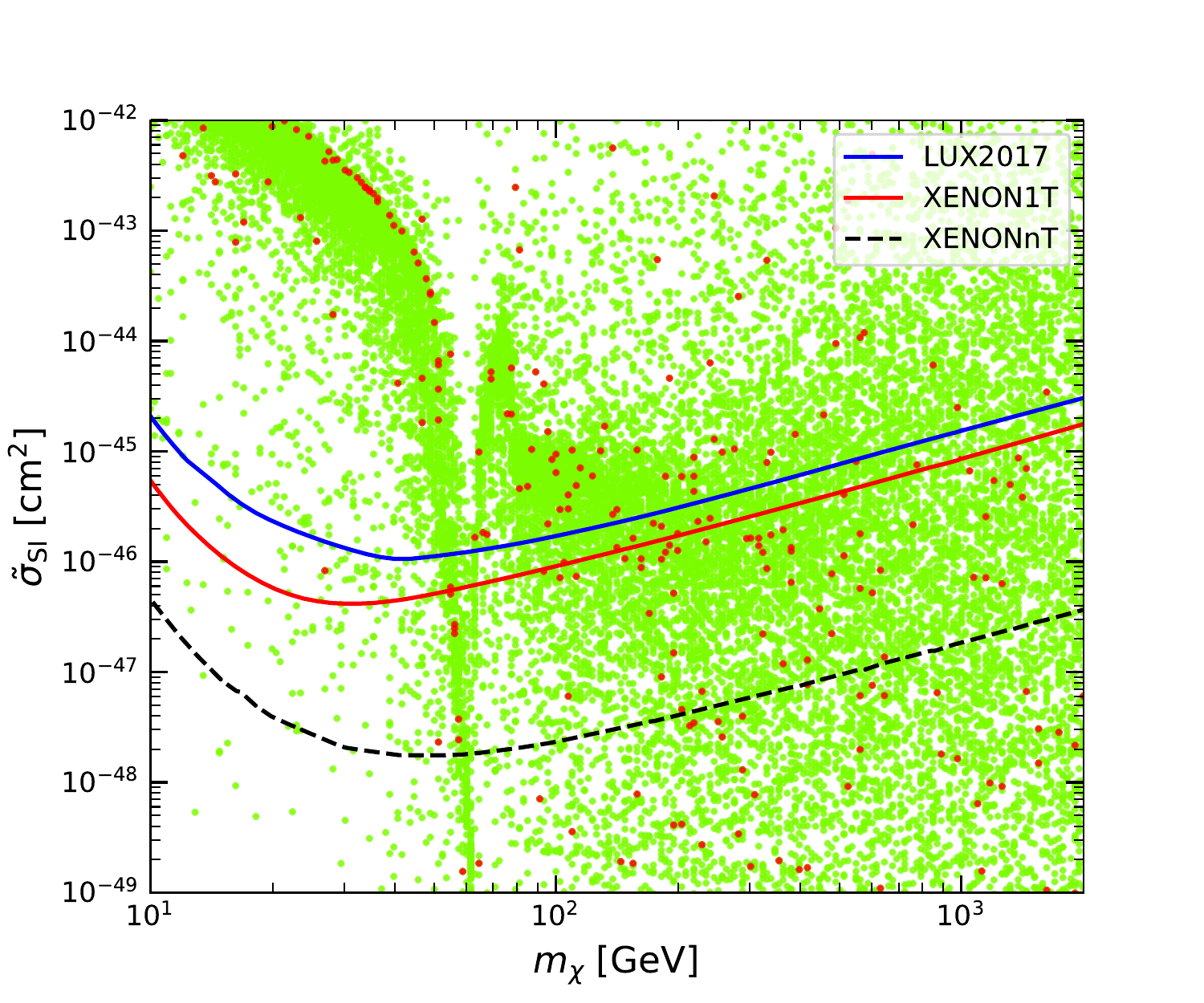}
    \caption{Effective SI DM-nucleon elastic scattering cross section as a function of the DM mass.
    The light green scatter points represent the samples that can trigger a type-II EWPT, and the red 
    scatter points are the samples further having a relic density in the range of $0.10-0.12$.}
    \label{fig:ptdd}
\end{figure}

In Fig.~\ref{fig:ptdd}, we plot our numerical scan result in the $m_{\chi}$-$\tilde{\sigma}_{\rm SI}$ plane in order to show the importance of the DM direct detection constraints on the parameter space of this $CP$-symmetric complex scalar model. 
The light green points represent the samples that can generate a sufficiently strong type-II EWPT, and the red points are those predicting a DM relic density in the range of $\Omega_{\rm DM} h^2 = 0.10-0.12$, 
close to the value of observed DM relic density from the Planck experiment.
From this plot, we see that there still exists a large parameter space in the model that have predictions below the most stringent constraint from DM direct detections.
As interpreted above, this is because the DM scattering with the nucleon in the direct detection via $\mathcal{H}$ and $\mathcal{S}$ portals are
suppressed by the mixing angle $\theta$, while the observed DM relic density can be achieved via the annihilation process 
$\chi\chi\to \mathcal{S}\mathcal{S}$, in which the coupling $\lambda_a$ plays the leading role.

\subsection{Final remarks}

\begin{table}[tbp]
    \renewcommand\arraystretch{1.5}
    \centering
    \caption{\label{tab:ii} A summary of parameters for model G1 and G2.}
    \begin{tabular}{|c|c|c|c|c|c|}
    \hline
    Model & $m_{\mathcal{S}}$ {\rm [GeV]}& $\theta$ & $\kappa_s$ & $c_{\chi}$ {\rm [GeV]} \\
    \hline
    G1 & 95.5 & $-0.086$ & 0.068 & $-501.2$ \\
    G2 & 141.3 & 0.234 & 0.191 & $-871.0$ \\
    \hline
    \end{tabular}
\end{table}
\begin{table}[tbp]
    \renewcommand\arraystretch{1.5}
    \centering
    \caption{\label{tab:iii} A summary of phase transition parameters for model G1 and G2.}
    \begin{tabular}{|c|c|c|c|c|c|}
    \hline
    Model & $T_c$ {\rm [GeV]}& $T_n$ [GeV] & $v/T_c$ & $\alpha$ & $\beta/H_n$ \\
    \hline
    G1 & 113.6 & 65.3 & 1.24 & 0.23 & 1044.2  \\
    G2 & 105.7 & 57.3 & 1.51 & 0.39 & 849.6 \\
    \hline
    \end{tabular}
\end{table}

Baryogenesis and gravitational waves resulted from the electroweak phase transition are two attractive topics that, however, are beyond the scope of this work.  Nevertheless, here we show two concrete $CP$-symmetric models, G1 and G2, in which laudable gravitational waves could be produced.  We fix the following parameters:
\begin{equation}
    w=500~{\rm GeV},~~m_{\chi}=300~{\rm GeV},~~{\rm and}~~\lambda_a=0.2
    ~.
\end{equation}
The other parameters are given in Table~\ref{tab:ii}.
We follow Ref.~\cite{Chiang2020JHEP} to calculate the bubble nucleation temperature $T_n$ and the gravitational wave parameters $\alpha$ and $\beta/H_n$ at the nucleation temperature.  A summary of the results for models G1 and G2 are given in Table~\ref{tab:iii}.
In figure~12 of Ref.~\cite{Chiang2020JHEP}, we show that the gravitational wave parameter region of $\alpha\gtrsim 0.1$ and $\beta/H_n\lesssim 10^4$ are within the sensitivity of LISA experiment \cite{{LISA2017}}, which thus could provide another test of our model.
The BBO interferometer~\cite{Crowder2005PRD} could further detect the gravitational wave parameter regions that are beyond the ability of LISA experiment.
In Table~\ref{tab:iii}, we also show the values of $v/T_c$, which is an important parameter for successful baryogenesis.
It has been shown that, for successful baryogenesis, the first-order EWPT should be strong enough so that the sphaleron process in the broken phase is sufficiently suppressed to avoid baryon asymmetry washout~\cite{Cline2000}. This leads to the conventional criterion $v/T_c\gtrsim 1$, which is satisfied in both models G1 and G2 (note that the phase transition parameter scan performed in section~\ref{sec:PTscan} has been restricted by this criterion). 
We leave the extensive study on the sphaleron process and gravitational waves in the $CP$-symmetric models for a future work.

\section{Summary}
\label{sec:summary}

In this work, we focus on studying the first-order type-II EWPT and the related DM phenomenology in the  real and complex singlet extensions of SM, amended with $\mathbb{Z}_2$ and $CP$ symmetries, respectively. 
Note that the type-II EWPT is attractive because it can provide a natural mechanism to generate a sufficiently strong first-order EWPT, a necessary condition for successful electroweak baryogenesis.
We first study the real singlet extension of SM with a $\mathbb{Z}_2$ symmetry.
We notice that the real singlet becomes a DM candidate after the first-order type-II EWPT. By scanning the entire parameter space, we have found that the model samples with the type-II EWPT suffer severely from the constraints of DM direct detection experiments by LUX and 
XENON1T. As a result, only a negligible fraction of ${\cal O}(10^{-4}-10^{-5})$ of DM relic density can be explained by the real singlet scalar.

We then turn to the complex singlet scalar extension of the SM with a {\it CP} symmetry, where we still examine the interplay between the first-order type-II EWPT and DM physics. 
In this model, due to the protection of the $CP$ symmetry, the imaginary component of $S$ can become a DM candidate. On the other hand, its real component can mix with the remaining real component in the Higgs doublet to form two physical states, ${\cal H}$ and ${\cal S}$, with the former much similar to the SM Higgs boson and the latter more singlet-like. We first perform a large-scale scan over the parameter space of 
physical interest, and find that there are ample parameter samples that can induce sufficiently strong type-II EWPT while satisfying the Higgs signal strength measurements at the LHC.  For the selected model parameters, we then explore their DM phenomenology, 
including the DM relic density and Higgs invisible decays, as well as DM direct and indirect detections. It turns out that there are still a 
large number of samples which can survive all of these DM constraints. In particular, we find parameters that can generate both strong type-II 
EWPT and correct DM relic abundance without any conflict with the current experimental data.  This clearly shows that the complex scalar model with the $CP$ symmetry is superior to the real scalar model in this regard.  Especially, when the DM $\chi$ is heavier than the mediator $\mathcal{S}$, the dominance of DM annihilation channel $\chi\chi \to \mathcal{S} \mathcal{S}$ during the DM freeze-out can help avoid the strong upper bounds of DM direct detections, since the two processes involve different parameter dependences, which is impossible for the real scalar model.

This $CP$-symmetric complex singlet model also predicts some new physics phenomena that can be further tested and constrained by future experiments. The mixing between the Higgs doublet field and the singlet field can lead to deviations in the cubic and quartic Higgs couplings, 
which in turn can significantly modify the di-Higgs production at colliders. In fact, our large-scale scan over the parameter space has already given some hint to the triple-Higgs coupling $\lambda_h v_0$ in this model with $\lambda_h$ restricted in the range of $0.1-0.3$, as shown in Fig.~\ref{fig:cpdist}.
Also, the mixing angle could be further restricted by the Higgs signal strength measurements and the DM direct detections. 
Moreover, the pair annihilation of the pseudoscalar DM $\chi$ may give rise to observable anomalies in the cosmic rays or $\gamma$-ray spectra. 
Finally, the strong EWPT can potentially generate a significant stochastic gravitational wave background, which can be probed by the future space-based gravitational wave detectors, such as LISA~\cite{LISA2017}, BBO~\cite{Crowder2005PRD}, and Taiji~\cite{Hu2017NSR,Ruan2020NA}.

\acknowledgments{This work was supported in part by the Ministry of Science and Technology (MOST) of Taiwan under Grant 
Nos.~MOST-108-2112-M-002-005-MY3, 108-2811-M-002-548 and 109-2811-M-002-550. 
DH is supported in part by the Chinese Academy of Sciences (CAS) Hundred-Talent Program and by National Science 
Foundation of China (NSFC) under Grant No. 12005254.
}

\bigskip

{\it Note added:} While preparing this manuscript, we found that the $CP$ symmetric model was also considered in a recent work by T.~Alanne {\it et al.,}~\cite{Alanne2020}.  Our work emphasizes on the type-II EWPT and the related DM phenomenology, while their study focuses on the type-I phase transition and the possible gravitational wave signals.

\end{document}